\documentclass[11pt]{article}

\usepackage{lineno}
\usepackage{natbib}
\usepackage{amssymb}
\usepackage{algorithm, algpseudocode}
\usepackage{algcompatible}
\usepackage{amsmath}
\usepackage{amsthm}
\usepackage{epsfig}
\usepackage{booktabs}
\usepackage{graphicx}
\usepackage{graphics}
\usepackage{float}
\usepackage{multirow}
\usepackage{color}
\usepackage{caption}
\usepackage{subcaption}
\usepackage{fullpage}
\usepackage[normalem]{ulem} 
\usepackage{makeidx}
\usepackage{xspace}
\usepackage{wrapfig}
\usepackage{tikz}
\usepackage{hyperref}
\hypersetup{
	colorlinks=true,
	citecolor=blue,
	linkcolor=blue,
	filecolor=blue,      
	urlcolor=blue,
}
\usepackage{setspace}

\newtheorem{proposition}{Proposition}
\newtheorem{assumption}{Assumption}
\newtheorem{Definition}{Definition}

\usepackage{dsfont}

\usepackage[affil-it]{authblk}

\begin{document}
	\begin{titlepage}
		\title{Modeling Network Interference with Multi-valued Treatments: the Causal Effect of Immigration Policies on Crime Rates
			\makeatother}
		\author[1]{Costanza Tort\'u}
		\author[1]{Irene Crimaldi}
		\author[3]{Fabrizia Mealli}
		\author[2]{Laura Forastiere}
		
		\affil[1]{IMT School for Advanced Studies Lucca}
		\affil[2]{Yale University}
		\affil[3]{University of Florence}
		\date{This version: May 2020.}
		\maketitle
		\vspace{-1cm}
		\begin{abstract}
			Policy evaluation studies, which intend to assess the effect of an intervention, face some statistical challenges: in real-world settings treatments are not randomly assigned and the analysis might be further complicated by the presence of interference between units. Researchers have started to develop novel methods that allow to manage spillover mechanisms in observational studies; recent works focus  primarily on binary treatments. However, many policy evaluation studies deal with more complex interventions. For instance, in political science, evaluating the impact of policies implemented by administrative entities often implies a multivariate approach, as a policy towards a specific issue operates at many different levels and can be defined along a number of dimensions. In this work, we extend the statistical framework about causal inference under network interference in observational studies, allowing for a multi-valued individual treatment and an interference structure shaped by a weighted network. The estimation strategy is based on a joint multiple generalized propensity score and allows one to estimate direct effects, controlling for both individual and network covariates. We follow the proposed methodology to analyze the impact of the national immigration policy on the crime rate. We define a multi-valued characterization of political attitudes towards migrants and we assume that the extent to which each country can be influenced by another country is modeled by an appropriate indicator, summarizing their cultural and geographical proximity. Results suggest that implementing a highly restrictive immigration policy leads to an increase of the crime rate and the estimated effects is larger if we take into account interference from other countries.
			\noindent \\
			\vspace{0.5cm}\\
			\noindent\textbf{Keywords:} causal inference; interference; complex network; multi-valued treatment; multiple network exposure; immigration policy\\
			
			\bigskip
		\end{abstract}
		\setcounter{page}{0}
		\thispagestyle{empty}
	\end{titlepage}

	\pagebreak \newpage
	

	\maketitle
	
	\linespread{1.5}\selectfont
	\section{Introduction \label{intro}}
	\subsection{Motivation}
	Policy evaluation studies aim to assess the effect of an intervention. Social sciences such as economics or political science often evaluate complex interventions, which have not been randomly assigned in the population. In some real-world settings, the analysis can be further complicated by the presence of interference between units. This phenomenon occurs because both economic and social agents are interconnected. Firms are connected by a wide mixture of juridical or commercial relationships including trading links, ownership or control ties and strategic alliances \citep{reinert2009princeton}. On the other side, individuals also interact through various mechanisms involving friendship or parental links, working collaborations or informative communications. In addition, even political entities are linked by way of explicit or velled agreements, or according to their specific geographical and cultural collocation with respect to a reference environment. These relations are depicted by a \emph{network}: the observed nodes are the elements of the population of interest, while network links represent the relations between them. Causal inference on a population of agents who are connected through a network faces some statistical challenges, including how to take into account the spillover mechanism that may arise. The typical causal inference framework \citep{rubin1980randomization} relies on a key assumption, called \emph{Stable Unit Treatment Value Assumption (SUTVA)}, which rules out the presence of interference among units. However, if agents are linked, experiments as well as observational studies may be affected by the presence of \emph{interference}, which formally occurs when the treatment of one unit has an effect on the response of other units (in addition to the unit's own outcome) \citep{cox1958planning}. In the presence of interference, the causal effect of a treatment  on one unit may be altered by the treatment received by other interfering units. For example, incentives targeted to some firms or companies may also benefit all those firms that are linked to them, according to juridical or economic relationships. In addition, policies implemented by single administrative entities also affect the outcomes of interfering territories. Dealing with interference is of paramount importance: wrongly assuming SUTVA can introduce a significant bias in the estimates and, consequently, lead to deceptive conclusions about the real effect of an intervention.
	\subsection{Related Works}
	For this reason, in recent years, a growing community of statisticians has started reasoning about interconnected units, developing novel methods and techniques which allow to account for interference in causal inference studies. The existing works extend the standard framework to include the network information in the definition of individual potential outcomes. Most of these works examine interference in \emph{randomized trials}. The limitations of the present statistical tools in dealing with possible dependencies among units have been first pointed out by \cite{rosenbaum2007interference}, who has also developed non-parametric tests to evaluate treatment and spillover effects in the presence of interference. This last study was extended, after a few years, by \cite{aronow2012general}, who presented a novel method to detect interference, and by \cite{bowers2013reasoning}, who proposed tools to model various dependency scenarios, also showing how to test hypotheses about causal effects according to the specific model that is supposed to depict interference. Recently, \cite{aronow2017estimating} rearranged the Horwitz-Thompson estimator allowing for the presence of interference  to obtain unbiased estimators for all the effects of interest, main and spillovers. \cite{athey2018exact} computed exact p-values for a variety of sharp null hypotheses about treatment effect in an experimental design where units are connected in an observed network. Interference may even play a role in the \emph{design of experiments}. Having proved that wrongly assuming SUTVA leads to biased results, \cite{eckles2017design} formalized a model of experiments in networks, proposing novel techniques for reducing this bias directly through the experimental design itself. Some other works focus on a particular type of interference known as \emph{partial (clustered) interference}, where units belong to exogenous groups and the spillover mechanism can occur only within clusters. The term "partial" is used here to counterpoise this scheme of clustered dependencies with the "general" interference scenario, where units interact according to a network. The partial interference assumption was formally introduced by \cite{sobel2006randomized} and it was further advanced also by \cite{hudgens2008toward}, who investigated the role of interference in the spreading of infectious diseases, where the probability that a person becomes infected is lower if the proportion of vaccinated individuals in his group is high \citep{basse2018analyzing}. Moreover, \cite{barkley2017causal} addressed the issue of a possible treatment selection among connected individuals and proposed causal estimands allowing for clustered dependence in the treatment selection \citep{papadogeorgou2019causal}. There are just a few articles that explicitly deal with general interference in \emph{observational studies}. For instance, \cite{hong2006evaluating} evaluated the policy of retaining low-achieving children in kindergarten rather than promoting them to first grade, using a multilevel propensity score model. \cite{van2014causal} and \cite{sofrygin2017semi} proposed a targeted minimum loss-based estimation (TMLE) estimator. A propensity score approach under spillover effects was first introduced by \cite{forastiere2016identification}, who presented a reworked formalization of the standard propensity score, named joint propensity score (JPS), with the aim of estimating the dose-response function in presence of interference. This work analyzes a binary treatment and models interference through an observed binary network. This last framework was employed by \cite{del2019causal} to explore trade distortions in agricultural markets: here the authors in turn rearranged the JPS formulation in order to model a continuous individual treatment.
	\subsection{Contribution}
	The existing statistical literature tackling interference in observational studies deals with binary or continuous treatments only. However, many policy evaluation studies involve more complex treatments, as, for example, treatments which are defined over more than two categories, known as multi-valued treatments. \emph{Multi-valued treatments} are highly diffused in nature. They are commonly used when the empirical aim consists in comparing various characterizations of an intervention and, above all, they are particularly employed in studies yearning to get the picture of complex and many-faceted phenomena, which may vary across multiple dimensions. For instance, evaluating the impact of different political attitudes towards puzzling macro-themes (immigration, national healthcare, economy) often calls for a multi-valued approach and requires also to  account for interference, since the treatment may spill over to different political entities. Since our empirical attempt is to evaluate the impact of immigration policy, we expand the theoretical framework proposed by \cite{forastiere2016identification} to the case of an individual multi-valued treatment, in observational studies. Generalization of the standard techniques (such as subclassification and propensity score methods) for binary treatments to the multi-valued scenario is not straight-forward and requires additional assumptions (\cite{lopez2017estimation}; \cite{yang2016propensity}; \cite{linden2016estimating}). The methodological approach becomes even more complicated if we decide to allow for the presence of interference, relaxing SUTVA and allowing for first-order spillover effects. The key idea is that under a multi-valued treatment, in the presence of interference, each unit is individually assigned to a treatment level and, simultaneously, they can be exposed to all the treatment levels, due to the interaction with their neighbors. Therefore, units experiment a \emph{multiple neighborhood exposure}, where each of their neighbors contributes in increasing the exposure to his own individual treatment level. In addition, the multiple neighborhood exposure mapping accounts for weights, which quantify the extent of dependencies, if they are observable. Weighted networks are widely spread in real-world data. For instance, networks of transactions between entities are usually enriched by the information about transactions' amount, social networks sometimes are coupled with the strength of friendship between units, scientific collaborations networks often provide the number of collaborations, political networks frequently measure the strength of connections between administrative and political entities by specific indicators. In settings with multi-valued treatments and weighted network, each unit is exposed to an individual treatment, which is categorical with a given number of categories, and to a neighborhood treatment, which is a multivariate continuous variable that measures the unit exposure to all treatment levels, resulting from the interaction of their neighbors and given the strength of these interactions.  Since we move in an observational study setting, where neither the individual treatment nor the neighborhood treatment are randomly assigned in the population, we propose an estimation strategy based on the usage of an extended version of the joint propensity score proposed by \cite{forastiere2016identification}. Our definition of propensity score, that we call \emph{Joint Multiple Generalized Propensity Score (JMGPS)}, allows to handle a multi-valued treatment and a multiple neighborhood exposure. The JMGPS is a type of generalized propensity score \citep{hirano2004propensity}, where the estimation strategy relies on a three-stage approach: i) we first assume a parametric distribution for both the individual and the neighborhood treatment and for the outcome variable; ii) for all the possible values that the joint treatment can assume, we use these models to predict missing potential outcomes; iii) we estimate the effects of interest comparing potential outcomes, and use bootstrap to compute the estimated standard errors.
	\\
	
	We make use of this methodology for the analysis of the causal effect of \emph{immigration policies} on \emph{crime rates}. In the last decades, the relevance of the immigration process has rapidly grown and shew the way to the spreading of a wide and open debate about the effects of migration. Some political parties, single politicians and citizens all around the world do believe that immigration represents a risk for national identity and, moreover, that it leads to a lower public safety. Consequently, they support governments that implement restrictive immigration policies. However, the causal effect of immigration policies on crime or social conditions in general has not been tested yet. In particular, there are not quantitative studies that involve and compare many countries, over a wide time frame. We analyze policies using the \emph{IMPIC} (Immigration Policies in Comparison) dataset that numerically measures all the immigration policies that have been implemented in the OECD countries from 1980 and 2010  in terms of restrictiveness. We include in the analysis 22 OECD countries that are located in Europe over the whole time frame covered by the IMPIC dataset. Our purpose is to investigate the impact of a national towards migrants on the \emph{crime rate}. In this application, the treatment of interest represents the restrictiveness of immigration policies, which is measured in the IMPIC Dataset through the evaluation of a series of single policies. Each policy refers to \emph{regulations} or \emph{control} protocols. The former are all the binding legal provisions that create or constrain rights (\cite{helbling2017measuring}), while the latter refer to the directives that have been adopted with the aim of monitoring whether the regulations are observed. Therefore, by aggregating items referring to these two political dimensions, we obtain two indicators measuring the country-year restrictiveness towards migrants, with respect to regulations and control mechanisms separately. Using this information, we define a multi-valued treatment by looking at the joint value of the two indicators, for each country-year profile. In this empirical setting, SUTVA is unlikely to hold. The political strategy towards migrants that a single country decides to implement may also affect crime rate of other countries. The possible spillover effect of the adopted political attitude towards immigration arises because migrants try to avoid countries with highly restrictive laws, and tend to move to states that appear to be more welcoming. Since migrants tend to move to countries with specific characteristics of their choice, the extent to which each country is affected by other countries' policies depends on their level of similarity.  Following this intuition, we derive an indicator summarizing the main factors which may prompt the spillover mechanism. These factors refer to various measures of similarity, which we reasonably believe to be the primary mechanisms driving interference. Specifically, this index, that we call \emph{Influence Index} (II), gives a measure of potential interference between each pair of countries at a given year and combines information about \emph{geographical proximity} and  \emph{cultural similarity}, which in turn are summarized by specific indicators.
	\\
	This work is organized as follows. In Section \ref{methodology} we focus on methodology. We first summarize the existing causal inference framework under interference in observational studies and then we present our methodological novelties: we introduce a multi-valued treatment and we propose a novel tool that allows one to model the neighborhood exposure in the presence of multi-valued treatment and weighted interference. We introduce the joint generalized multiple propensity score and we illustrate the estimation strategy. In Section \ref{application} we motivate the importance of the empirical application, giving a broad overview of the existing literature and briefly describing data. Moreover, we give a more detailed characterization of the Influence Index and we provide a deeper explanation on the definition of treatment nominal categories. In Section \ref{results} we present the main empirical results. Then, in the appendix, we collect the proofs of the theoretical propositions we present in Section \ref{methodology}, we give the precise definition of II also adding further details about the definition of the neighborhood treatment variable, and we present the detailed results of all the models we implement, reporting some descriptives and checking the robustness of the main findings with respect to alternative definitions of the treatment variable. 
	
	\section{Methodology \label{methodology}}
	
	In this section we explain the main methodological developments. We start from the existing causal inference framework under interference and then we present the novel approach for multi-valued treatments (Subsection \ref{causnetint}). Finally, we define the Joint Multiple Generalized Propensity Score (Subsection \ref{jmgps}) together with its properties and we propose the estimation strategy (Subsection \ref{estimproc}). 
	\subsection{Causal Inference Under Network Interference}
	\label{causnetint}
	The main scope of causal inference is estimating the effect of a treatment on some outcome variable in a population of units. Let us consider a sample $\mathcal{N}$ composed of $N$ units. Denote as $K$ the number of treatment levels and let  $Z_i \in \{1, \dots, K \}$ be a categorical variable representing the treatment assigned to unit $i$ and $Y^{obs}_i$ the observed outcome for the same unit. By $\textbf{Z}$ and $\textbf{Y}^{obs}$ we denote the corresponding vectors of the whole sample $\mathcal{N}$. Moreover, $\textbf{X}_i$ denotes a vector of $P$ covariates (or pre-treatment variables) that are not influenced by the treatment assignment. Following  \cite{rubin1974estimating,rubin1980randomization}, we postulate, for each unit, the existence of $K$ potential outcomes, one for each treatment vector, $Y_{i}(\textbf{Z})$. Most causal inference relies on the Stable Units Treatment Assumption (SUTVA) \citep{rubin1986comment}. 
	SUTVA consists of two different components:  (i) the Individualistic Treatment Response (ITR) \citep{manski2013identification} (or  \textit{no interference}) assumption, which states that there is no interference between units, each unit's potential outcomes are defined only by the unit's own treatment; (ii) the \textit{consistency} (or \textit{no multiple version of the treatment}) assumption, which states that there are no different versions of the treatment levels. As a consequence, under SUTVA, potential outcomes can be indexed only by $Z_{i}$ -i.e $Y_{i}(Z_{i})$- and the observed outcome is the one corresponding to the treatment that each unit $i$ has actually received:  $Y_{i}^{obs}=Y_{i}(Z_{i})$.
	\\
	
\begin{figure}[H]
		\centering
		\begin{subfigure}{.5\textwidth}
			\centering
			\includegraphics[width=75mm]{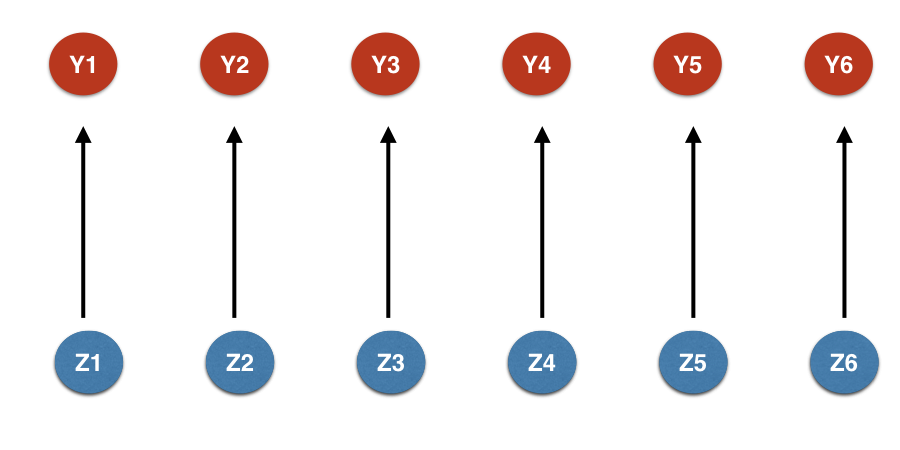}
			\caption{No interference scenario}
			\label{fig:nointerference}
		\end{subfigure}%
		\begin{subfigure}{.5\textwidth}
			\centering
			\includegraphics[width=75mm]{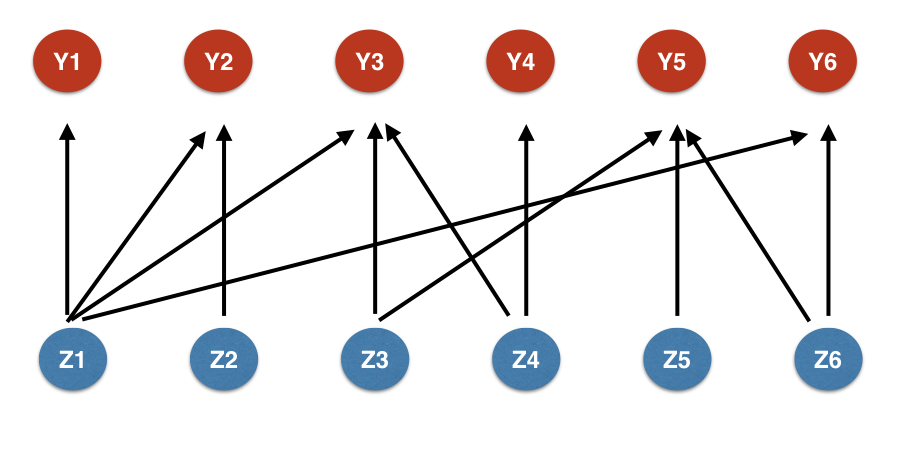}
			\caption{With Interference scenario}
			\label{fig:interference}
		\end{subfigure}
		\caption{No-Interference vs Interference scenarios: blue nodes represent individual treatments, red nodes their corresponding outcomes.	\label{fig:interferencecomp}}
	\end{figure}

\begin{figure}[H]
		\begin{center}
			\includegraphics[width=50mm]{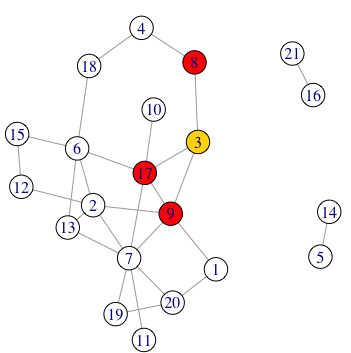}
			\caption{Neighborhood of a given node: the figure shows a given unit (yellow-colored unit) and highlights his own neighbors (red-colored units), in a population of connected agents.}
			\label{fig:neigh}
		\end{center}
	\end{figure}

	SUTVA completely rules out the presence of interference among units. However, in many real situations, this no-interference assumption is violated. This phenomenon can occur in various and heterogeneous frameworks. For instance, in economics, firms assigned to a program of incentives can be affected by incentives received by other firms. In epidemics, vaccines are known to benefit the whole community, including unprotected individuals, because they reduce the reservoir of infection and the infectiousness. Finally, in political sciences, policies implemented in some administrative regions may have an effect also on neighboring territories. All these examples refer to empirical situations in which one unit's outcome may be influenced by other units' treatment level. Figure \ref{fig:interferencecomp} gives a graphical intuition of interference. In the no-interference case, each unit's outcome (red nodes) is affected only by his own treatment (blue nodes); in the presence of interference, neighboring units' treatments also affect the individual outcome.
	When the spillover mechanism comes into play, wrongly assuming SUTVA leads to biased results and, consequently, to inaccurate or even misleading conclusions about the effects of interest. 
	In order to model interference, we must look at the relationships between units. We consider an observed undirected network $\mathcal{G}=(\mathcal{N},\mathbb{E})$, where $\mathcal{N}$ is the set of nodes (the population of interest) and $\mathbb{E}$ represents the set of edges indicating links between nodes. 
	For each node $i$, we identify a partition of $\mathcal{N}$ into two subsets: i) the neighborhood of node $i$, $\mathcal{N}_{i}$, that includes all the nodes $j$ with a link with node $i$, $i \leftrightarrow j$, and we denote by $N_i$ the cardinality of $\mathcal{N}_i$; ii) the  No-Neighborhood of node $i$, $\mathcal{N}_{-i}$, including all the nodes $j$ without a link with node $i$,  $i \nleftrightarrow j$. According to these partitions, we define, for each node $i$, the following partitions of the treatment vector and of the outcome vector, ($Z_{i},\textbf{Z}_{\mathcal{N}_{i}},\textbf{Z}_{\mathcal{N}_{-i}}$), ($Y_{i},\textbf{Y}_{\mathcal{N}_{i}},\textbf{Y}_{\mathcal{N}_{-i}}$). Figure \ref{fig:neigh} shows the neighborhood of a given node.
	Admitting network interference in the analysis implies the replacement of SUTVA by an assumption on the interference structure. \cite{forastiere2016identification} make a neighborhood interference assumption, which allows for the existence of first-order spillover effects between neighbors, in the context of binary treatments. More precisely, using the notation $Y_{i}(\boldsymbol{Z})$ for potential outcomes of unit $i$, we have: 
	\begin{assumption}[Stable Unit Treatment on Neighborhood Value Assumption (SUTNVA) \label{ass: sutnvabin}]
		SUTNVA is constituted by two components:
		\begin{enumerate}
			\item No Multiple Versions of Treatment (\emph{Consistency}):
			$Y_{i}(\textbf{Z})=Y_{i}(\textbf{Z}^{1}) \; \; \forall \textbf{Z}, \textbf{Z}^{1}$ such that $\textbf{Z}=\textbf{Z}^{1}$, that is, the mechanism used to assign the treatments does not matter.
			\item Neighborhood Interference:
			There exists a function $g_{i} : \{0, 1\}^{N_{i}} \to \mathbb{G}$, with $\mathbb{G} \underline{\subset} \mathbb{R}$ such that, for all $ \textbf{Z}_{\mathcal{N}_{-i}}, \textbf{Z}^{1}_{\mathcal{N}_{-i}}$ and $ \textbf{Z}_{\mathcal{N}_{i}},\textbf{Z}^{1}_{\mathcal{N}_{i}}$ with $g_{i}( \textbf{Z}_{\mathcal{N}_{i}})=g_{i}(\textbf{Z}^{1}_{\mathcal{N}_{i}})$, we have 
			$$Y_{i}(Z_{i},\textbf{Z}_{\mathcal{N}_{i}},\textbf{Z}_{\mathcal{N}_{-i}})=Y_{i}(Z_{i},\textbf{Z}^{1}_{\mathcal{N}_{i}},\textbf{Z}^{1}_{\mathcal{N}_{-i}}).$$ This assumption basically states that there is interference and it is modelled by the function $g_{i}$. 
		\end{enumerate}
	\end{assumption}	
	The variable $G_{i}=g_{i}(\boldsymbol{Z}_{\mathcal{N}_{i}})$, called \emph{neighborhood treatment}, represents the unit’s exposure to the treatment, due to the influence of his neighbors. The function $g_{i}$ can be defined in many different ways, according to the interference mechanism that is assumed to take place. For instance, it can simply count the number of treated neighbors or it can measure the proportion of treated neighbors.  Note that under SUTNVA interference is assumed to arise only from neighborhood of each unit and that any higher order interference is completely ruled out. This means that unit $i$ is not influenced by units other than their neighbors. This restriction over the interference structure may appear to be strong in some scenarios, but it seems to be plausible in many empirical applications. 
	\\
	
	In many real-world applications treatments are implicitly or explicitly multi-valued. In epidemics, researchers are interested in comparing between drugs \citep{linden2016estimating}. In economics, firms are exposed to different types of incentives. In training programs, participants receive different types of coaching \citep{cattaneo2010efficient}. Finally, political scientists evaluate political strategies towards highly complex and multi-faceted issues which involve different sub-fields. In such scenarios, a common practice is to collapse the multi-valued treatment into a binary variable, but this approach implies a relevant loss in terms of information and it prevents the possibility of capturing differential effects across treatment levels \citep{cattaneo2010efficient}. For this reason, researchers have started to study how to extract causal information under multi-valued treatments, developing novel assumptions and techniques that extend to the multi-valued scenario standard causal inference methods as matching, subclassification, inverse probability weighting on the propensity score \citep{lopez2017estimation}. However, no existing work suggests how to deal with interference in the multi-valued scenario. In this work we fill in this gap. \\
	
	Under first-order spillover effects, each unit is exposed to a neighborhood treatment, defined as a numerical synthesis neighbors' treatment status. In the binary setting, this synthesis is usually expressed by a single value, $G_{i} \in \mathbb{G}\underline{\subset}\mathbb{R}$, while when the individual treatment is defined by multiple categories this definition of $G_{i}$ is too simplistic. Categorical treatments imply a more complex definition of the neighborhood treatment exposure, as the neighborhood treatment must summarize the individual network exposure to \emph{each} treatment level. The mathematical tool that we introduce to model the neighborhood treatment under multi-valued individual treatment is the \emph{Neighborhood Treatment Exposure Matrix} $\boldsymbol{G}$:	
	\begin{Definition}[Neighborhood Treatment Exposure Matrix (NTEM), $\boldsymbol{G}$]
		\label{def:NTEM}The NTEM is an $N \times K$ matrix $\boldsymbol{G}$ that collects the unit neighborhood exposure to all the treatment levels:
		$$\boldsymbol{G}=\left(\begin{array}{ccccc} G_{1,1} & \dots & G_{1,z} & \dots & G_{1,K} \\ \vdots &\vdots  &\vdots  &  \vdots &  \vdots  \\ G_{i,1} & \dots & G_{i,z} & \dots & G_{i,K} \\ \vdots &\vdots  &\vdots  &  \vdots &  \vdots  \\G_{N,1} & \dots & G_{N,z} & \dots & G_{N,K} 
		\\
		\end{array}\right).$$
		Each element $G_{i,z} \in \mathbb{G}\underline{\subset}\mathbb{R}$ indicates the exposure of unit $i$ to the treatment level $z\in\{1,\dots,K\}$. Each row is the neighborhood treatment vector for the unit $i$, $\boldsymbol{G}_{i} \in \mathbb{G}^{K}\underline{\subset}\mathbb{R}^K$. Therefore, the neighborhood treatment is not a scalar measure, as in the binary treatment setting. It is instead a $K$-dimensional vector whose components describe the unit's neighborhood exposure to each treatment level. 
	\end{Definition}
	Hence, the first component of the recalled SUTNVA for binary treatments, i.e.~the no multiple versions of treatment assumption, is confirmed as stated above; while the second component, i.e.~the neighborhood interference assumption, is here replaced by a more general assumption, which handles the spillover mechanism generated by a multi-valued individual treatment:
	\begin{assumption}[Multiple Neighborhood Interference]
		There exists a function $g_{i} : \{1,\dots ,K\}^{N_{i}} \to \mathbb{G}^{K}$, with $\mathbb{G}^{K} \underline{\subset} \mathbb{R}^{K}$ , such that, for all $ \textbf{Z}_{\mathcal{N}_{-i}}, \textbf{Z}^{1}_{\mathcal{N}_{-i}}$ and $ \textbf{Z}_{\mathcal{N}_{i}},\textbf{Z}^{1}_{\mathcal{N}_{i}}$ with $g_{i}( \textbf{Z}_{\mathcal{N}_{i}})=g_{i}(\textbf{Z}^{1}_{\mathcal{N}_{i}})$, we have 
		$$Y_{i}(Z_{i},\textbf{Z}_{\mathcal{N}_{i}},\textbf{Z}_{\mathcal{N}_{-i}})=Y_{i}(Z_{i},\textbf{Z}^{1}_{\mathcal{N}_{i}},\textbf{Z}^{1}_{\mathcal{N}_{-i}}).$$ This assumption states that interference is modelled by the function $g_{i}$ (with components $g_{i,z}$, $z\in\{1,\dots,K\}$) which maps the neighborhood exposure of unit $i$ over a $K$-variate domain, that is $\boldsymbol{G}_{i}=g_{i}(\textbf{Z}_{\mathcal{N}_{i}})$.
	\end{assumption}
	If interference is modeled through a weighted network, the function $g_{i}$ must take into account the weights, $I_{ij}$, measuring the strength of the link between $i$ and the neighbor $j$. For instance, given an individual treatment with $K$ categories, we can set $G_{i,z}=\sum_{j \in \mathcal{N}_{i}} I_{ij}\delta_{zj}$,
	where $\delta_{zj} $ is a dummy variable that equals 1 if $Z_{j}=z$ and 0 otherwise. 
	\\
	
	Each unit $i$ is exposed to a \emph{joint treatment} $(Z_{i}, \boldsymbol G_{i})$: the \emph{individual treatment} $Z_{i}$, which is a categorical variable with $K$ levels and the \emph{neighborhood multi-treatment} $ \boldsymbol{G}_{i}$ which is a $K$-variate variable. Hence, potential outcomes, for each unit $i$, are indexed by the joint treatment: $Y_{i}(Z_{i},\boldsymbol G_{i})=Y_{i}(Z_{i}=z, \boldsymbol G_{i}=\boldsymbol{g})$. The observed outcome is the one corresponding to the actual joint treatment each unit is exposed to: $Y_{i}^{obs}=Y_{i}(Z_{i},\boldsymbol G_{i})$. 
	\\
	
	Regarding the effects of interest, the number of the possible comparisons is ${{K}\choose{2}}=\frac{K!}{(K-2)!2!}$. Under the multi-valued individual treatment, the direct effect of a given treatment $z'$ with respect to the treatment $z$, keeping the neighborhood treatment as fixed, can be expressed as 
	\begin{equation}
	\tau_{z'z}(\boldsymbol{g})=E \big[Y_{i}(z',\textbf{g})-Y_{i}(z,\textbf{g}) \big].
	\end{equation}
	This quantity represents the individual causal effect of a direct exposure, when the neighborhood treatment is set to $\boldsymbol{g}$. The overall main effect can be define averaging the individual treatment effect over the multivariate probability distribution of the neighborhood treatment, that is
	\begin{equation}
	\tau_{z'z}=\sum_{\boldsymbol{g}\in \mathbb{G}^{K}}\tau_{z'z}(\boldsymbol{g})P(\boldsymbol{G}_{i}=\boldsymbol{g}).
	\end{equation}
	
	\subsection{Joint Multiple Generalized Propensity Score (JMGPS)\label{jmgps}}
	In this work, we focus on observational studies, where neither the individual nor the neighborhood treatment are randomly assigned in the population. The general strategy in observational studies is to control for baseline covariates such that, conditioning on them, the treatment assignment becomes as good as random. In other words, we can exclude any dependence between treatment variable and potential outcomes. This assumption is known as \emph{unconfoundness} \citep{rosenbaum1983central}.
	In some empirical applications, the number and nature of covariates makes it hard to control for all of them without relying on strong parametric assumptions and, extrapolating in these settings, researchers, instead of conditioning on the set of covariates, prefer to work with a scalar synthesis of them, called \emph{propensity score} \citep{rosenbaum1983central}.  In the binary treatment setting with no interference, propensity score is defined as the conditional probability of receiving the treatment, given the values of the covariates. If the unconfoundness assumption is valid when conditioning on individual covariates, it remains valid when conditioning on the propensity score. Using this approach, researchers benefit from a relevant dimensionality reduction in the analysis. 
	This general approach, which is well grounded in the standard causal inference literature, can be extended to the setting with multi-valued treatment and interference. Here the unconfoundness assumption must be related to the joint treatment and the joint potential outcomes. 
	Following the motivations proposed by \cite{yang2016propensity}, we rely on the weaker version of unconfoundness with respect to the individual multi-valued treatment. Hence, instead of considering the actual multi-valued treatment variable $Z_{i}$, we refer to $K$ \emph{treatment indicator variables} representing the presence ( or absence ) of a given treatment level $z$, $D_i(z)$. Thus, we advance the following assumption:
	\begin{assumption}[Weak Unconfoundedness of the Joint Treatment\label{ass: weekunc}]
		$$
		P(D_i(z)=1,\,\boldsymbol{G}_i=\boldsymbol{g}\,|\,Y_i(z,\boldsymbol{g}), \boldsymbol{X}_i)
		=
		P(D_i(z)=1,\,\boldsymbol{G}_i=\boldsymbol{g}\,|\,\boldsymbol{X}_i)
		\; \; \forall z \in \{ 1, \dots, K \} \; \; \forall \boldsymbol{g} \in \mathbb{G}^{K}.
		$$
	\end{assumption}
	Note that, in presence of interference, $\boldsymbol{X}_{i}$ can include purely individual covariates as well as neighborhood covariates. From now on, we denote as $\textbf{X}_{i}^{ind}$ the individual covariates and as $\textbf{X}_{i}^{neigh}$ the neighborhood covariates.
	\\
	
	In the presence of interference, the propensity score is the joint probability of receiving a value $z$ of the individual treatment and, simultaneously, being exposed to a value $\boldsymbol{g}$ of the neighborhood treatment, given the unit's baseline covariates.  \cite{forastiere2016identification} formally introduced propensity score under network interference in the case of a binary treatment. We expand their definition allowing for a multi-valued individual treatment and a multivariate neighborhood treatment. Therefore, we introduce the \emph{Joint Multiple Generalized Propensity score (JMGPS)} as follows:
	\begin{Definition}[Joint Multiple Generalized Propensity score (JMGPS) \label{def:JMGPS}], $\psi(z,\boldsymbol{g},\boldsymbol{x})$
		\\
		JMGPS is the probability of being jointly exposed to a $K$-variate individual treatment equal to $z$ and to a $K$-dimensional neighborhood treatment equal to $\boldsymbol{g}$, conditioning on baseline covariates. 
		\begin{equation}
		\psi(z,\boldsymbol{g}; \boldsymbol{x})= P(Z_{i}=z,\boldsymbol{G}_{i}=\boldsymbol{g}|\textbf{X}_{i}=\textbf{x})
		\end{equation}
	\end{Definition}
	As the standard propensity score, the Joint Multiple Generalized Propensity Score is a balancing score, that is, it guarantees balance with respect to neighborhood and individual covariates. JMGPS has the following properties.
	\begin{proposition}[Balancing Property of JMGPS \label{prop:bal}]
		The joint propensity score is a balancing score, that is
		$$ P(Z_{i}=z, \boldsymbol{G}_{i}= \boldsymbol{g}|\boldsymbol{X}_{i})=P(Z_{i}=z, \boldsymbol{G}_{i}= \boldsymbol{g}|\psi(z,\boldsymbol{g};\boldsymbol{X}_{i})), \; \; \forall \; z \in \{1, \dots, K \} \; \text{and} \; \forall \; \boldsymbol{g} \in \mathbb{G}^{K}.
		$$
		Proof in Appendix \ref{Proof 1}
	\end{proposition}
	Furthermore, conditioning on JMGPS, we can exclude any dependency between the treatment variable and potential outcomes. 
	\begin{proposition}[Conditional Unconfoundedness of  $D_{i}(z)$  and $\boldsymbol{G}_{i}$ given JMGPS \label{prop:condunc}]
		Under Assumption  \ref{ass: weekunc}, for all $z \in \{1, \dots, K \} $ and  $ \boldsymbol{g} \in \mathbb{G}^{K}$
		$$
		P(D_i(z)=1,\,\boldsymbol{G}_i=\boldsymbol{g}\,|\,Y_i(z,\boldsymbol{g}), \psi(z,\boldsymbol{g};\boldsymbol{X}_i))
		=
		P(D_i(z)=1,\,\boldsymbol{G}_i=\boldsymbol{g}\,|\,\psi(z,\boldsymbol{g};\boldsymbol{X}_i)).
		$$
		\\
		Proof in Appendix \ref{Proof 2}
	\end{proposition}
	
	Following \cite{forastiere2016identification}, we rely on the factorization of the joint propensity score in neighborhood propensity score and individual propensity score. 
	\begin{Definition}[Factorization of the Joint Multiple Generalized Propensity score (JMGPS)]
		JMGPS can be factorized as follows
		\begin{equation}
		\begin{aligned}
		\psi(z,\boldsymbol{g}; \boldsymbol{x})&= P(Z_{i}=z,\boldsymbol{G}_{i}=\boldsymbol{g}|\textbf{X}_{i}=\textbf{x}) \\
		&  =P(\boldsymbol{G}_{i}=\boldsymbol{g}|Z_{i}=z,\textbf{X}_{i}^{g}=\boldsymbol{x}^{g})P(Z_{i}=z|\boldsymbol{X}_{i}^{z
		}=\boldsymbol{x}^{z}) \\
		&=\lambda(\boldsymbol{g}; z,\boldsymbol{x}^{g})\phi(z; \boldsymbol{x}^{z}),
		\end{aligned}	
		\end{equation}
		where $\lambda(\boldsymbol{g}; z,\boldsymbol{x}^{g})$ is the neighborhood propensity score and $\phi(z; \boldsymbol{x}^{z})$ is the individual propensity score. $\boldsymbol{X}_{i}^{z}$ and $\boldsymbol{X}_{i}^{g}$ are vectors collecting covariates that affect the individual and the neighborhood treatment, respectively. Note that the two sets corresponding to the covariates included in $\boldsymbol{X}_{i}^{z}$ and $\boldsymbol{X}_{i}^{g}$ may differ. In particular, $\textbf{X}_{i}^{g}$, can collect individual covariates as well as neighborhood covariates, while $\boldsymbol{X}_{i}^{z}$ includes individual variables only.
	\end{Definition}
	Using the factorization that we have just presented, we illustrate another key property of JMGPS.
	\begin{proposition}[Conditional Unconfoundedness of $D_{i}(z)$ and $\boldsymbol{G}_{i}$ given individual and neighborhood propensity scores ] \label{prop:conduncin}
		Under Assumption  \ref{ass: weekunc}, for all $z \in \{ 1, \dots, K \}$ and $\boldsymbol{g} \in \mathbb{G}^{K}$, we have 
		$$
		P(D_i(z)=1,\,\boldsymbol{G}_i=\boldsymbol{g}\,|\,Y_i(z,\boldsymbol{g}), \phi(z;\boldsymbol{X}^{z}_{i}),\lambda(\boldsymbol{g};z,\boldsymbol{X}_{i}))
		=
		P(D_i(z)=1,\,\boldsymbol{G}_i=\boldsymbol{g}\,|\, \phi(z;\boldsymbol{X}^{z}_{i}),\lambda(\boldsymbol{g};z,\boldsymbol{X}_{i})).
		$$
		Proof in Appendix \ref{Proof 3}
	\end{proposition}
	This property indicates that conditioning on the two components separately still guarantees the validity of the conditional unconfoundness property. 
	\subsection{Estimation Procedure \label{estimproc}}
	The JMGPS is the fundamental element of the estimation procedure that we propose here in this section. Both its components can be seen as peculiar characterizations of the generalized propensity score proposed by \cite{hirano2004propensity}. This procedure follows a parametric approach and imputes missing potential outcomes for all configurations of the joint treatment and then compares them to estimate the direct effects of interest. Standard errors and confidence intervals are computed using bootstrap methods. The proposed estimation strategy can be summarized in three main steps.
	\\
	
	\noindent
	\textbf{1) Model treatment and outcome variables } \\
	(1.a) Assume a distribution for $Z_{i}$, $\boldsymbol{G}_{i}$ and $Y_{i}$. \\
	Formally:
	$$Z_{i}\sim f^{z}(\boldsymbol{X}^{z}_{i};\theta^{z}),$$
	$$\boldsymbol{G}_{i}\sim f^{g}(Z_{i},\boldsymbol{X}^{g}_{i};\theta^{g}),$$
	$$Y_{i}(z,\boldsymbol{g}) \sim f^{y}(z,\boldsymbol{g},\phi(z;\boldsymbol{X}_{i}),
	\lambda(\boldsymbol{g};z,\boldsymbol{X}_{i});\theta^{y} ).$$
	Of course, the multi-valued characterization of $Z_{i}$ demands for the definition of a statistical model for categorical responses, with respect to the individual propensity score model. Furthermore, $\boldsymbol{G}_{i}$ requires the definition of a multivariate model. \\
	(1.b) Predict actual individual and neighborhood propensity score \\
	Estimate the parameters $\theta^{z}$ and  $\theta^{g}$  of the models for $Z_i$ and $\boldsymbol{G}_i$;
	Use the estimated parameters in Step 1,  $\widehat{\theta^{z}}$ and $\widehat{\theta^{g}}$,  to predict for each unit $i \in \mathcal{N}$ the actual individual propensity score and the actual neighborhood propensity score, that is, the probabilities of being exposed to the individual treatment and the multivariate neighborhood treatment they have actually being exposed to:
	$$\widehat{\Phi_{i}}=\phi(Z_{i};\boldsymbol{X}^{z}_{i};\widehat{\theta}^{z}),$$
	$$\widehat{\Lambda_{i}}=\lambda(\boldsymbol{G}_{i}; Z_{i},\boldsymbol{X}_{i}^{g};\widehat{\theta}^{g}).$$
	(1.c) Estimate parameters of the outcome model \\
	Use the predicted propensity scores $\widehat{\Phi}_{i}$ and $\widehat{\Lambda}_{i}$, in order to estimate the parameters $\theta^{y}$ of the outcome model $Y_{i}(z,\boldsymbol{g})$:
	$$Y_{i}\sim f^y(Z_{i},\boldsymbol{G}_{i},\widehat{\Phi}_{i},\widehat{\Lambda}_{i};\theta^{y}).$$
	\textbf{2) Impute Missing Potential Outcomes
	} \\
	Consider the domain of the joint treatment $(Z_{i} = z, \boldsymbol{G}_{i} = \boldsymbol{g})$. In particular, $\boldsymbol{G}_{i}$ is a $K$-dimensional continuous variable.
	For each possible value of the joint treatment, that is, for each combination $(Z_{i} = z, \boldsymbol{G}_{i} = \boldsymbol{g}) \; \text{s.t} \; z \in \{1, \dots, K \},\; \boldsymbol{g} \in \Gamma$ \footnote{In order to explore a multivariate domain, one common practice is constructing a $K$-dimensional discrete grid that scours the possible values of $\boldsymbol{g}$, over its $K$ components' respective domain. Let us denote this grid as $\Gamma$, $\Gamma \subset \mathbb{G}^{K}$.}, with $\Gamma \subset \mathbb{G}^{K}$, and for each unit $i \in \mathcal{N}$,\\
	(2.a) Predict the individual propensity score corresponding to that level of $z$, $\widehat{\phi}(z;\boldsymbol{X}^{z}_{i})$.\\
	(2.b) Predict the neighborhood propensity score corresponding to that level of $\boldsymbol{g}$, $\widehat{\lambda}(\boldsymbol{g}; z,\boldsymbol{X}^{g}_{i})$.\\
	(2.c) Use that estimated parameters to impute the potential outcome $\widehat{Y}_{i}(z,\boldsymbol{g})$, that is,
	$$\widehat{Y}_{i}(z,\boldsymbol{g}) \sim
	f^y(z,\boldsymbol{g},\widehat{\phi}(z;\boldsymbol{X}^{z}_{i}),\widehat{\lambda}(\boldsymbol{g}; z,\boldsymbol{X}^{g}_{i});\widehat{\theta}^{y}).$$
	\noindent		
	\textbf{3) Estimate the effects of interest and their corresponding variance}
	\\
	(3.a) Estimate the final direct effects of interest, averaging potential outcomes over $\lambda(\boldsymbol{g}; z,\boldsymbol{X}^{g}_{i})$:
	$$\widehat\tau_{z'z}=\frac{1}{N}\sum_{i=1}^{N}\Big[  \sum_{\boldsymbol{g} \in \Gamma} \big(\widehat{Y}_{i}(z',\boldsymbol{g}) -  \widehat{Y}_{i}(z,\boldsymbol{g})\big) \Big]  \frac{\widehat{\lambda}(\boldsymbol{g}; z,\boldsymbol{X}^{g}_{i})}{\sum_{\boldsymbol{g'} \in \Gamma} \widehat{\lambda}(\boldsymbol{g'}; z,\boldsymbol{X}^{g}_{i})}.$$
	\\
	(3.b) Compute variance through bootstrap. This procedure works as follows. Choose a large value $R$. For $R$ times, $r:\{1,\dots,R\}$, draw a random sample $\mathcal{N}_{r}$ with replacement from $\mathcal{N}$. Estimate the effects of interest over the subsample $\mathcal{N}_{r}$, so getting $\widehat\tau_{z'z}(r)$. Consider the distribution of $\widehat\tau_{z'z}(r)$ over the $R$ repetitions $\widehat{T}^{r}_{z'z}$. The estimated standard error $\text{Std.Er}(\widehat\tau_{z'z})$ is the standard error of $\widehat{T}^{r}_{z'z}$.

\section{Empirical Application 	\label{application}}
In this section, we focus on the empirical application. We first explain the relevance of our empirical research question with respect to the existing literature about immigration (Subsection \ref{rq}). Second, we describe the different data sources that we have merged (Subsection \ref{data}). Finally, we formalize and discuss the influence index (Subsection \ref{ici}) and we explain how we derive the treatment categories (Subsection \ref{trcat}) also showing how to implement the estimation strategy we have presented in the previous section (Subsection \ref{jmgpsestim}). 
\subsection{Empirical Research Question \label{rq}}

In the last decades, interest about immigration has rapidly grown, so that it has become a major topic both in academic and real life debates (\cite{helbling2017measuring}). Immigration flows significantly increased, since many people attempted to move away from countries which have been suffering long periods of wars and bad economic conditions. The consequence of this process is that the world has become multicultural: migrants have started to be socially included into the hosting countries, searching for a new job or even getting married. Moreover, migrants have diffused their own social and religious beliefs. However, immigration has entailed not only positive outcomes. Some countries which have embraced a relevant number of migrants, have experienced the rising of social tensions (\cite{rudolph2003security}). Over the last decades, economic conditions gradually get worse: unemployment rates rose up, and real wages went down. In addition, people have noticed a relevant worsening in the perception of individual security. The conviction that immigration may have aggravated these negative processes has slowly taken root in the public opinion. Politicians and common citizens have started to evaluate problematic consequences about immigration and globalization. Concerns about migration spread up in three main directions. First, native people perceive immigration as a risk for the preservation of national identity. Integration results to be not always easy and multiculturalism tends to be perceived more as a threat than as an opportunity. Second, as migrants move looking for better living conditions, they represent, in the common belief, competitive profiles for job. Finally, people tend to blame migrants for raising crime \citep{bigo2002security}. 
\\
In recent years, many researchers have started studying the effects of the increasing migration flows. For instance, \cite{bove2016does} has assessed the effect of migration on the diffusion of terrorism, while \cite{rudolph2006national} has evaluated the effects on national security. 
Many epidemiological studies as \cite{polissar1980effect}, \cite{stillman2007migration} and \cite{hildebrandt2005effects} have analyzed the consequences on the spreading of some diseases and on public health, in general.  Furthermore, \cite{coleman2008demographic} and \cite{keely2000demography} have studied the causal effect of migration on some demographic outcomes. \cite{bianchi2008immigration} \cite{bianchi2012immigrants} and \cite{stansfield2016reevaluating} have studied the impact of flows on crime. The existing works assessing the causal link between migration flows and crime present findings which are conceptually in contrast to common perception, suggesting that increasing immigration flows does not lead to higher crime rates. Some of them also state that there is actually a negative effect of immigration on crime. 
\\

The public discussion about migration has also involved the immigration policies that national governments  implement with the aim of controlling and ruling the immigration process. \cite{brochmann1999mechanisms} defined immigration policies as the \textit{"government’s statements of what it intends to do or not do (including laws, regulations, decisions or orders) in regards to the selection, admission, settlement and deportation of foreign citizens residing in the country"}.
These policies can be more or less restrictive and, therefore, can discourage or encourage migrants, respectively. Some political parties of various countries all over the world support the idea that implementing restrictive immigration policies limits the negative effects of migrations and, consequently, leads to better living and economic conditions for the natives. On the other side, many politicians and intellectuals argue that the legislative system of a country should encourage immigrants and facilitate their settling. 
\\

\begin{figure}[H]
	\centering
	\includegraphics[width=60mm,height=60mm]{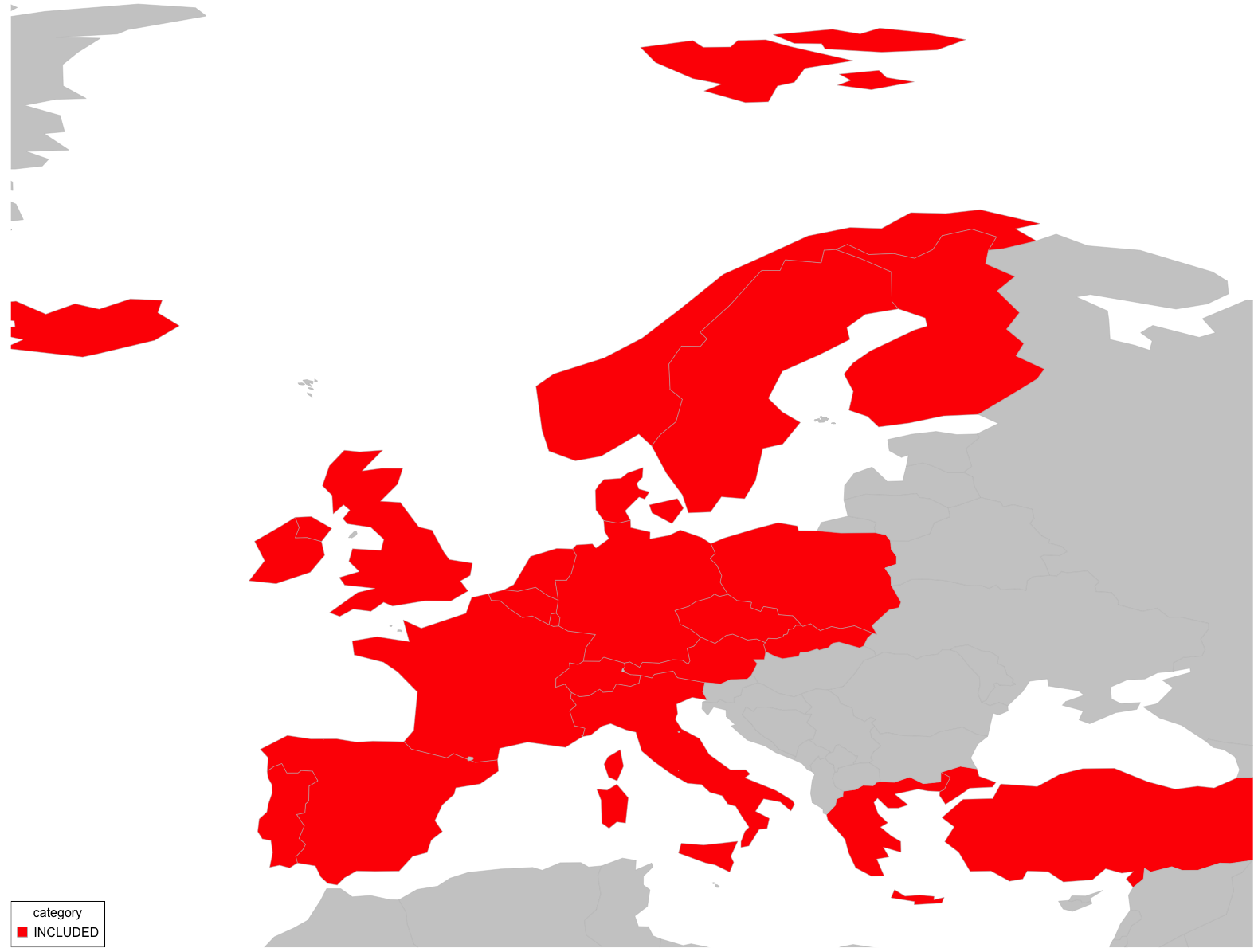}
	\caption{Included countries: red colored countries are the ones included in our analysis. \label{fig:included}}
\end{figure}

In this work, we investigate the causal effect of immigration policies on crime rates. Specifically, we study the effect of the restrictiveness of the implemented immigration policy on one year lagged national \emph{crime rate}, expressed in terms of homicides every 10.000 inhabitants.  We approach this research question from a country level perspective: in particular, we focus on the subset of OECD countries that are located in the Continental Europe and we inspect their policies towards migrants from 1980 and 2010. These policies have been measured in terms of restrictiveness in the \emph{IMPIC (Immigration Policies in Comparison)} Dataset (\cite{helbling2017measuring}, \cite{schmid2016validating}), that properly conceptualizes and quantitatively compares many national policies that affect migrants \citep{munck2002conceptualizing}.  Starting from the restrictiveness measures supplied by the IMPIC Dataset and taking into account the conceptualization of the observed policies that the same dataset proposes, we evaluate the national immigration policy over two political dimensions: restrictiveness of regulations and restrictiveness of control strategies. From now on, we denote by \emph{Reg} and \emph{Cont} the two variables representing those two dimensions.
We present a treatment variable that qualitatively distinguishes country-year profiles with respect to these two measures and we pairwise compare different political strategies.
\\
Our empirical analysis covers 22 OECD countries that are situated in the continental Europe \footnote{24 OECD countries are located in the Continental Europe but we remove from the analysis Hungary and Estonia as they present extreme values of the crime rate.}. These countries are characterized by very different immigration experiences: there are countries that have experienced increasing immigration since one or two centuries  (Great Britain, Germany, France), countries that recently turned from emigration to immigration (Italy, Spain) and countries that have experienced very limited immigration (Finland) \citep{helbling2017measuring}. However, they are still highly comparable from an institutional point of view, as they are all fully developed democracies and are all located in Europe. In Figure  \ref{fig:included} we show a map of the 22 countries included. 
We believe that this is an innovative contribution to the existing literature about migration. Indeed, there are just some recent studies (\cite{geddes2016politics}, \cite{messina2007logics}) about immigration policies but they look at particular behaviors of individual countries or describe a small number of countries \citep{helbling2017measuring}.  Even if there are existing works that assess the effects of migration flows on crime rates, they focus on single countries comparing subnational administrative entities and they do not take into account the national political strategy towards migrants. Moreover, they all rule out spillover effects.
\\

In this empirical scenario, interference may play a relevant role. Migrants may choose to avoid highly restrictive countries and to settle in places where laws appear to be more welcoming. But we expect that they try to preserve some characteristics of their settling choice. Thus, the general idea is that dependence between two countries is related to their level of similarity. We assume that two mechanisms may drive interference: \emph{geographical proximity} and  \emph{cultural proximity}. Thus, we build a continuous indicator that analytically captures these driving mechanisms of interference. Each component contributes according to a given weight. We test various configurations of the influence weights in order to check the robustness of our results with respect to different restrictions about dependencies.  

\subsection{Data \label{data}}
This work merges different data sources. First, we use the \emph{IMPIC (Immigration Policies in Comparison)} Dataset  \citep{helbling2017measuring} that provides information about national immigration policies. In particular, this dataset includes data on migration policies for all the OECD countries over thirty years (from 1980 to 2010). Policies are measured with respect to their restrictiveness, from 0 (less restrictive) to 1 (highly restrictive). Data include more than 50 policies for each country-year profile and items are aggregated in different indicators of the general restrictiveness towards migrants with respect to the \emph{regulation} and \emph{control} protocols. The former aspect is related to all the laws that discipline immigrants and their life in the hosting country, while the latter is referred to the mechanisms that help in monitoring whether the regulations are abided by \citep{schmid2016validating}.
\\

Second, we handle different datasets to assemble the \emph{Influence Index} (II), which measures the extent of dependency between each pair of countries at a given year, that is the extent to which a country's immigration policy influences the crime rate of another country. As we will fully discuss in the forthcoming section, this index is a convex combination of two complex indicators quantifying \emph{geographical proximity} and \emph{cultural similarity} between two countries at a given year. We build up the \emph{geographical proximity indicator} starting from the CEPII Dist Dataset (\cite{mayer2011notes}) which includes different measures of bilateral distances (in kilometers) and a dummy variable denoting pairwise contiguity. Furthermore, we explore \emph{cultural similarity} between each pair of countries at a given year looking at the \emph{linguistic similarity} through the CEPII Language Dataset \citep{melitz2014native} and at the \emph{religious similarity} through CEPII Gravity Dataset \citep{fouquin2016two}. 
\\

Third, we make use of some datasets that provide country-year features. Specifically, we collect information about crime rates relying on the \emph{World Countries Homicide rate dataset} which comprises information about the country-year specific number of homicides per 10.000 inhabitants. In addition, we manage the \emph{World Development Indicators} dataset, provided by the World Bank (\cite{coppedge2018v}, \cite{lindberg2014v} and \cite{coppedge2018c}) which contains highly detailed country-year indicators referring to various aspects of society: they quantitatively mark out the economic situation, the demographic features, the state of the social welfare and democracy and even the level of equality, freedom and justice. 
\\

The observed population is characterized by country-year observations: we deal with $C=22$ countries observed over $T=30$ years where the initial time $t=1$ is year 1980 while the ending time $t=T$ is year 2010).  \footnote{Estonia starts to be included in the analysis from 1991, after its Independence. Czech Republic and Slovak are instead considered only from 1993: they both became independent countries after the Dissolution of Czechoslovakia which took effect on Jan 1, 1993.}. Therefore, the generic unit $i$ is a pair $(c,t)$ and the total number of units is $N=C \times T$. We indicate as $\textbf{Y}^{obs}=\{Y^{obs}_{ct}\}$ the $(N \times 1)$ observed (country,year) outcome vector. 
Furthermore, we take into account of the pre-treatment covariates matrix $\textbf{X}$ with dimension $N \times P $ : each row of this matrix represents a country-year observation, while each column refers to a specific baseline factor. The included covariates can be grouped in four sets, according to the main issue they refer to: i) \emph{Economy}: GDP per capita, equal distribution of resources index, state ownership of economy index; ii) \emph{Inequality}: educational inequality index, income inequality index, health equality index, power distributed to gender index, equal access; iii) \emph{Freedom and Participation}: civil participation index, freedom of expression index, freedom of religion; iv) \emph{Demography}: life expectancy, fertility rate. We denote as $\mathcal{X}$ the set collecting these variables.
\\

\begin{figure}[H]
	\centering
	\includegraphics[width=100mm]{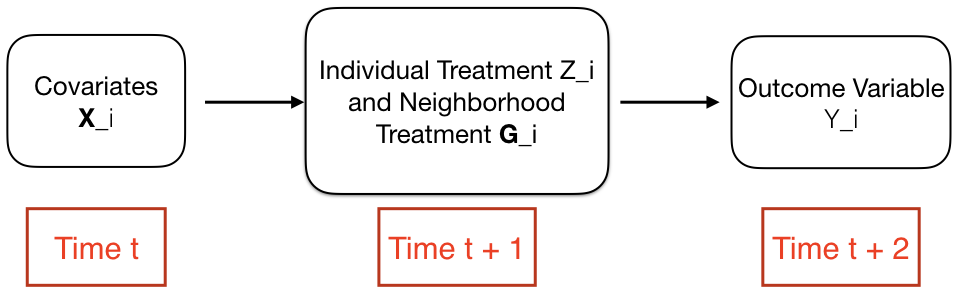}
	\caption{Variables Effects: timing 	\label{fig: timingvar}}
\end{figure}

We point out that we assume that there is a \emph{one-year lag effect} of baseline covariates on treatment and of treatment on outcome variables. We state that the covariates of one country $c$  at time $t$ affects his individual as well as neighborhood treatment at time $t+1$ and that the joint treatment in turn affects the outcomes at time $t+2$. Figure \ref{fig: timingvar} provides an intuition of this conceptual idea.
There is no way to test this assumption, but a lagged process of causation seems plausible in the considered empirical scenario. In addition, in order to avoid reverse causality issues, in the propensity score estimation we control for no-lagged outcome variables. For instance, we consider the baseline covariates of one country at time $t$ to model his joint treatment at time $t+1$ and consequently his outcome at time $t+2$ and in the set of the pre-treatment variables at time $t$ we include the outcome at time $t$ as well. 
\\	

\subsection{Modelling Interference: Influence Index (I) \label{ici}}
Here, we must define the interference structure taking into account the possible mechanisms that could make immigration policies in one country affecting the crime levels of other countries. The idea is that immigrants avoid highly restrictive countries and settle to areas that are similar to the first choice with respect to some characteristics, but more politically welcoming. Thus, the relevance of spillover between each pair of countries depends on their pairwise similarity. We assume that the kind of similarity that plays a role in this mechanism is the geographic proximity ( meaning, the geographic distance between countries ) and the cultural similarity. In some sense, we state that a migrant, who is willing to move, chooses the most welcoming alternative among the countries that are relatively near and culturally similar to the first choice option, that though implements highly restrictive laws. 
Therefore, we build up a composite indicator which numerically summarizes these two mechanisms which we reasonably believe are the key prompters of dependency. The two components contribute to the determination of the global index according to some weights, $\alpha$ and $\beta$. This index, that we call \emph{Influence Index (I)}, gives a unique information about how much one country $c$ interfere with a country $c'$ at time $t$. Formally:
\begin{Definition}[Influence Index (I)]
	\label{def:ici}
	$$I_{cc',t}=\alpha \times IG_{cc'}+\beta \times IC_{cc',t}$$
	where $IG_{cc'} \in [0,1]$ is the \emph{geographic proximity indicator} which measures the geographical proximity between country $c$ and country $c'$ and $IC_{cc',t} \in [0,1]$ is the \emph{cultural similarity indicator} that measures the cultural similarity between two countries time $t$. The constants $\alpha$ and $\beta$, with $\alpha + \beta = 1$, are the Influence Inputs Weights (IIW) that determine the extent to which each component contributes to the global index.
\end{Definition}
Note that, since the Influence Index is a convex combination of two indicators bounded between 0 and 1, it is in turn bounded between 0 and 1, that is, $ I_{cc',t} \in [0,1]$. More details about the construction of the Influence index can be found in the Appendix \ref{iciapp}. We test various allocations of $IIW$ to check the robustness of our results with respect to different assumptions over the interference structure. Following the same approach of many existing works in economics and social sciences (\cite{del2019causal}), we have ruled out the presence of intertemporal links, that is we set $I_{(ct),(c't')}=0 \;\;  \forall c,c',t,t' \;\text{with}\; t \ne t'$.  

\subsection{Treatment Categories \label{trcat}}

\begin{figure}[H]
	\centering
	\begin{subfigure}{.48\textwidth}
		\centering
		\includegraphics[width=75mm]{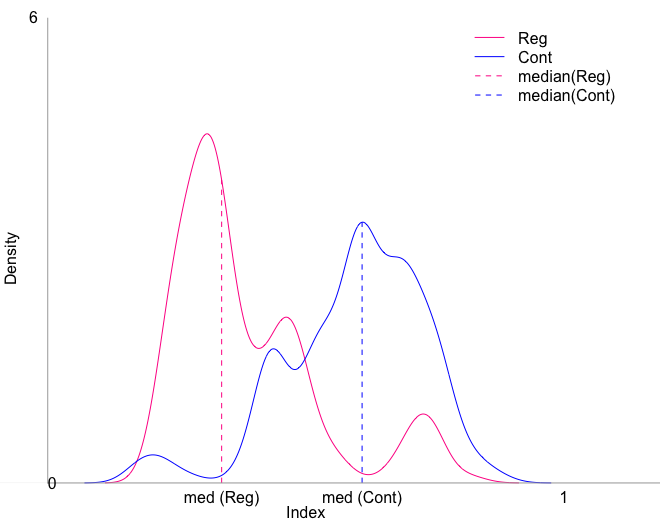}
		\vspace{+0.3cm}
		\caption{Density distributions of \emph{regulation} (violet line) and \emph{control} (blue line) indexes, and their respective medians (dotted lines) 	\label{fig: treatcat_dens}}
	\end{subfigure}\hspace{0.08 cm}
	\begin{subfigure}{.48\textwidth}
		\centering
		\vspace{-0.05cm}
		\includegraphics[width=85mm]{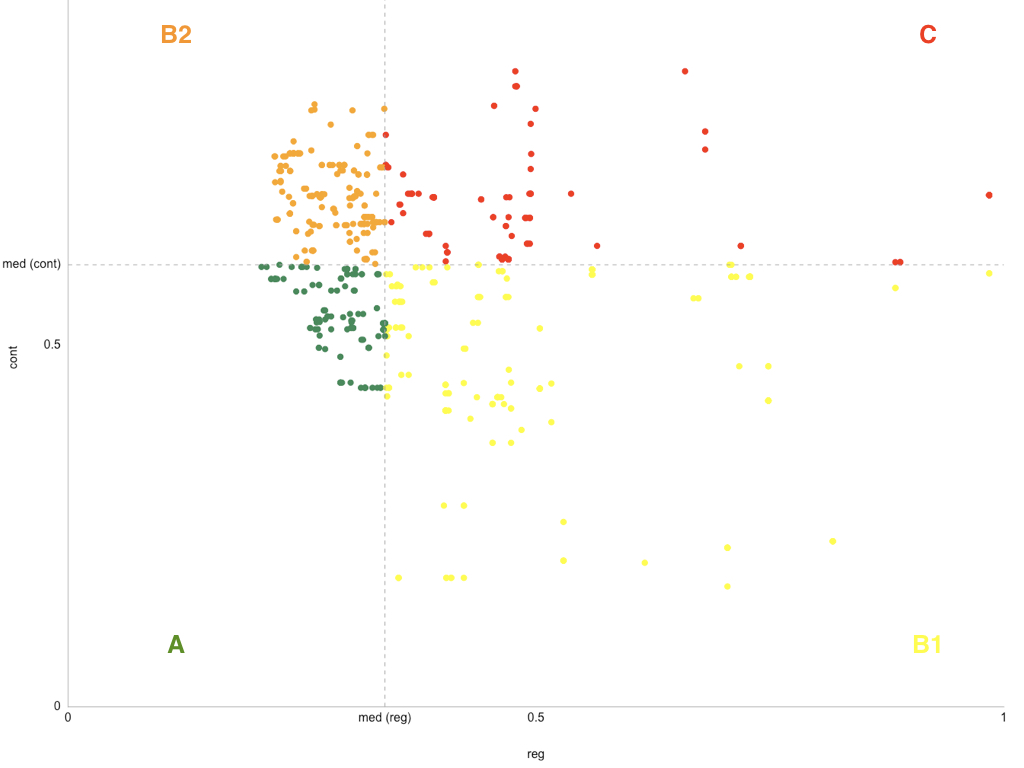}
		\caption{Individual collocation in the treatment categories according to the regulation and control indexes	\label{fig: treatcat_colloc}}
	\end{subfigure}
	\caption{Treatment Categories Definition	\label{fig: treatcat}}
\end{figure}

IMPIC dataset provides indicators which measure the country-year restrictiveness towards migrants with respect to \emph{regulations} and \emph{control} mechanisms. Let us denote as $reg_{i}$ the reported value of the restrictiveness in terms of regulations of the generic country-year profile $i=(c,t)$ and with $cont_{i}$ the corresponding value in terms of control. 
We define the nominal treatment categories looking at the joint distribution of the two indicators. In particular, denoting as $\text{med}_{Reg}$ and $\text{med}_{Cont}$ the median of the distribution of the regulations indicator and of the control one, respectively, we define the treatment categories as follows
\begin{Definition}[Nominal Treatment Categories	\label{def: ind tr}] Individual treatment is obtained by applying the following categorization criterion.
	\begin{itemize}
		\item $Z_{i}$=LL if $\text{reg}_{i} \le \text{med}_{Reg}$ and $\text{cont}_{i} \le \text{med}_{Cont}$: this category identifies profiles that are barely restrictive with respect to the two mechanisms. 
		\item $Z_{i}$=HL if $\text{reg}_{i} > \text{med}_{Reg}$ and $\text{cont}_{i} \le \text{med}_{Cont}$: this category detects profiles which implement restrictive regulations but weak control strategies.
		\item $Z_{i}$=LH if $\text{reg}_{i} \le \text{med}_{Reg}$ and $\text{cont}_{i} > \text{med}_{Cont}$: this category indicates a welcoming attitude in terms of regulations but intense control protocols. 
		\item $Z_{i}$=HH if $\text{reg}_{i} \ge \text{med}_{Reg}$ and $\text{cont}_{i} \ge \text{med}_{Cont}$: this category denotes an highly restrictive policy towards migrants with respect to both regulations and control.
	\end{itemize}
\end{Definition}
Figure \ref{fig: treatcat} provides a graphical idea of the previously described definition procedure. The left subfigure shows the density distributions of the \emph{regulation} and \emph{control} indexes: their corresponding median values are identified by dotted lines while black colored line represents the underlying distribution of the whole immigration policy index which results by a weighted mean of the former two \citep{helbling2017measuring}. The right subfigure shows the individual treatment collocation based on their own values of the \emph{regulation} and \emph{control} indexes.

\begin{figure}[H]
	\centering
	\begin{subfigure}{.48\textwidth}
		\centering
		\includegraphics[width=82mm]{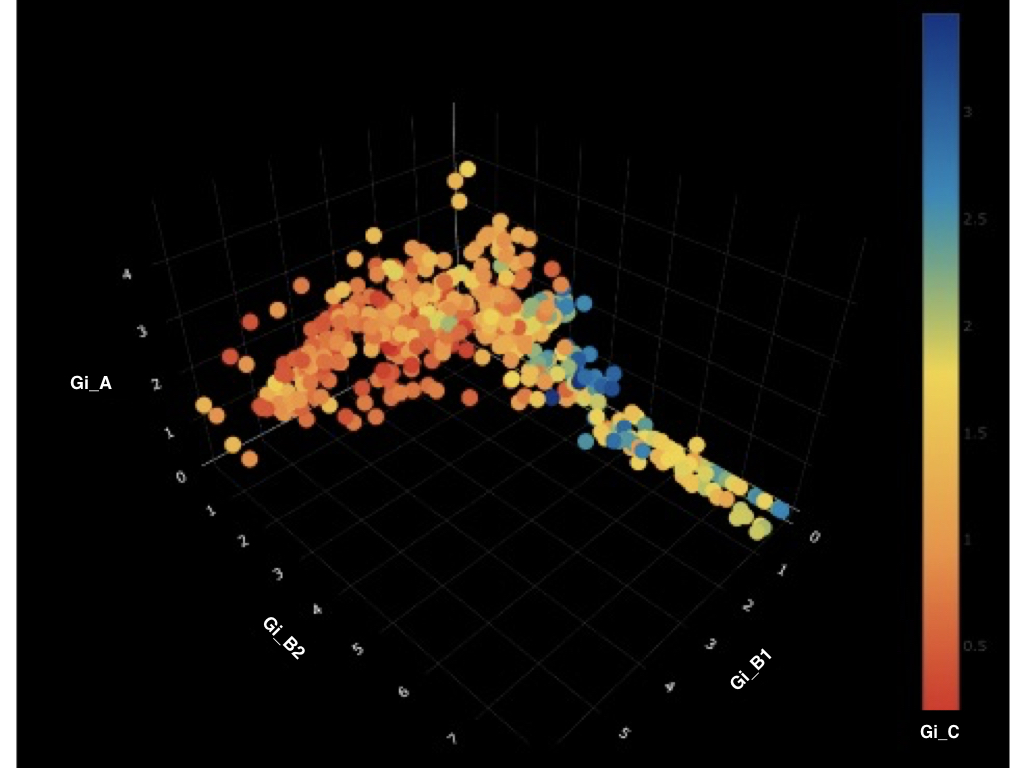}
		\caption{Tridimensional scatterplot of the neighborhood treatment variable \label{fig: neightr_3dsc} }
	\end{subfigure}\hspace{0.08 cm}
	\begin{subfigure}{.48\textwidth}
		\centering
		\includegraphics[width=88mm]{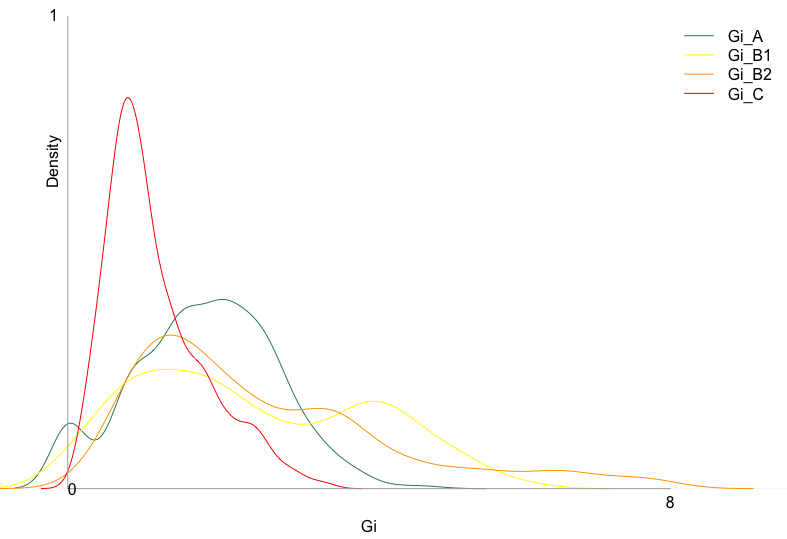}
		\caption{Density distribution of the singular components that constitutes $\boldsymbol{G}_{ct}$ \label{fig: neightr_dens}}
	\end{subfigure}
	\caption{Neighborhood Treatment, $\alpha=\beta=\frac{1}{2}$	\label{fig: neightr}}
\end{figure}

Hence, we deal with a $K$-valued individual treatment, where $K=4$. 
Let us  denote as $\textbf{Z}=\{Z_{ct}\}$, the $(N \times 1)$ multi-valued treatment vector where $Z_{ct} \in \{ LL, HL, LH, HH \}$. Following Definition \ref{def: ind tr} and assuming the Influence Index as the ruling mechanism of dependencies,  we explicit the neighborhood treatment $\boldsymbol{G}_{ct}$ as 
$$\boldsymbol{G}_{ct}=\left(\begin{array}{c} G_{ctLL} \\ G_{ctHL}\\  G_{ctLH}\\ G_{ctHH}   \\
\end{array}\right)= \left(\begin{array}{c} \sum_{c' \in \mathcal{N}_{ct}} I_{cc',t}\delta_{Ac't} \\ \sum_{c' \in \mathcal{N}_{ct}} I_{cc',t}\delta_{HLc't} \\  \sum_{c' \in \mathcal{N}_{ct}} I_{cc',t}\delta_{LHc't}  \\
\sum_{c' \in \mathcal{N}_{ct}} I_{cc',t}\delta_{HHc't}  \\
\end{array}\right),$$
where $\delta_{LLc't}, \delta_{HLc't}, \delta_{LHc't},,\delta_{HHc't}$ are dummy variables such that $\delta_{LLc't}=1$ if $Z_{c',t}=LL$ and 0 otherwise; $\delta_{HLc't}=1$ if $Z_{c',t}=HL$ and 0 otherwise; $\delta_{LHc't}=1$ if $Z_{c',t}=LH$ and 0 otherwise; $\delta_{HHc't}=1$ if $Z_{c't}=HH$ and 0 otherwise.  Consequently, the potential outcomes are defined as $Y_{ct}(Z_{ct}, \boldsymbol G_{ct})$.
Figure \ref{fig: neightr} displays the distribution of the neighborhood treatment variable under the hypothesis of equal contribution to the Influence index of the cultural and geographical subcomponents,  ($\alpha=\beta=\frac{1}{2}$). 

In order to estimate the causal effects of interest, we follow the estimation procedure described in Section \ref{methodology}. 

\subsection{Joint Multiple Generalized Propensity Score (JMGPS) Estimation \label{jmgpsestim}}
We estimate the two components of JMGPS (see Definition \ref{def:JMGPS}), that is the individual propensity score and the neighborhood propensity score. 
\subsubsection{Individual Propensity Score \label{indpsestim}}
The individual propensity score $\phi(z;\boldsymbol{x}^{z})$ is the individual probability of receiving an individual treatment $z$ conditioning on unit-level baseline covariates.
If the individual treatment is a categorical variable with $K$ nominal categories the estimation strategy consists in fitting a model for categorical responses. Here we use the \emph{Multinomial Logit Model} (\cite{agresti2018introduction}; \cite{long2006regression} and \cite{menard2002applied}), where the reference category is set to "LL", that is,

$$P(Z_{i}=LL)=\frac{1}{1+\sum_{z\ne LL}\exp{\boldsymbol{\beta}_{z} \boldsymbol{X}_{i}^{z}}},$$
$$P(Z_{i}=HL)=\frac{\exp{\boldsymbol{\beta}_{HL} \boldsymbol{X}_{i}^{z}}}{1+\sum_{z\ne LL}\exp{\boldsymbol{\beta}_{z} \boldsymbol{X}_{i}^{z}}},$$
$$P(Z_{i}=LH)=\frac{\exp{\boldsymbol{\beta}_{LH} \boldsymbol{X}_{i}^{z}}}{1+\sum_{z\ne LL}\exp{\boldsymbol{\beta}_{z} \boldsymbol{X}_{i}^{z}}},$$
$$P(Z_{i}=LH)=\frac{\exp{\boldsymbol{\beta}_{HH} \boldsymbol{X}_{i}^{z}}}{1+\sum_{z\ne LL}\exp{\boldsymbol{\beta}_{z} \boldsymbol{X}_{i}^{z}}}.$$

\begin{figure}[H]
	\centering
	\begin{subfigure}{.48\textwidth}
		\centering
		\includegraphics[width=80mm]{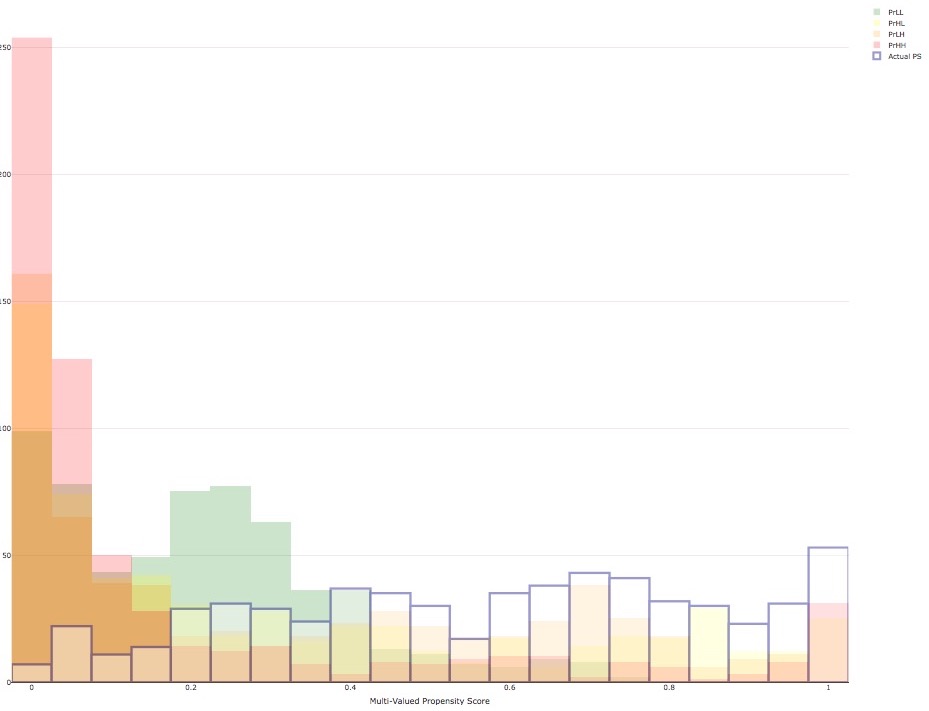}
		\caption{Histograms of $\phi(LL;\boldsymbol{X}^{z}_{i};\widehat{\theta}^{LL})$ (green), $\phi(HL;\boldsymbol{X}^{z}_{i};\widehat{\theta}^{HL})$(yellow), $\phi(LH;\boldsymbol{X}^{z}_{i};\widehat{\theta}^{LH})$(orange), $\phi(HH;\boldsymbol{X}^{z}_{i};\widehat{\theta}^{HH})$ (red) and $\phi(Z_{i};\boldsymbol{X}^{z}_{i};\widehat{\theta}^{z})$  \label{fig: indpsh}}
	\end{subfigure}\hspace{0.08 cm}
	\begin{subfigure}{.48\textwidth}
		\centering
		\includegraphics[width=81mm]{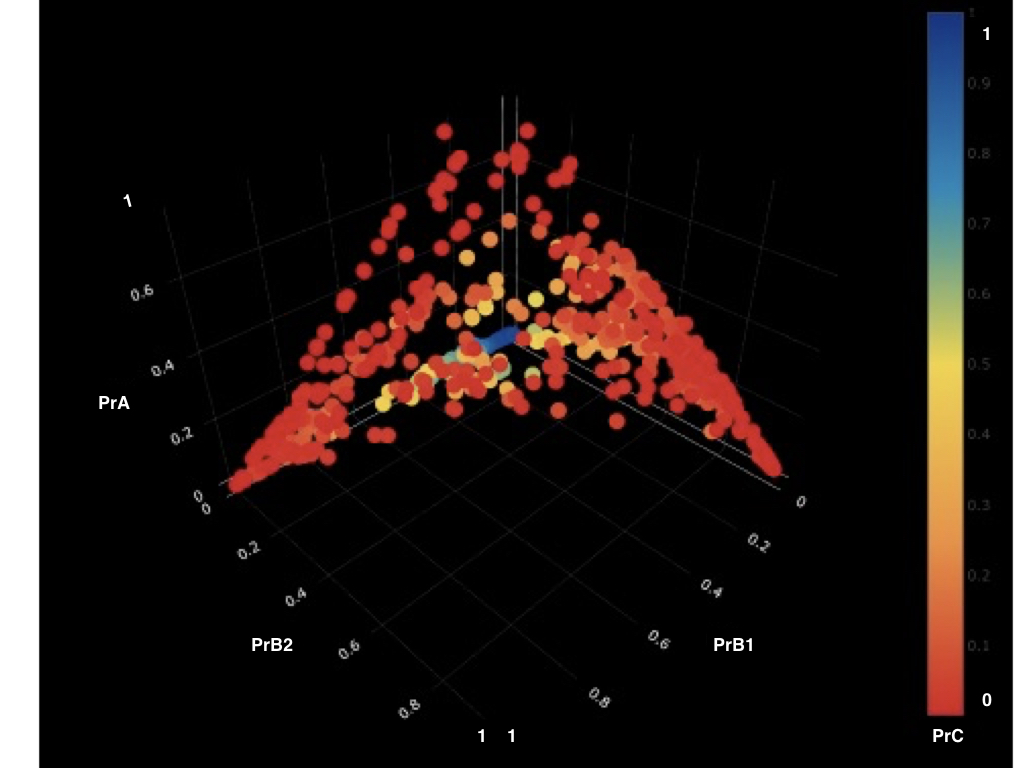}
		\caption{Tridimensional scatterplot of $\phi(LL;\boldsymbol{X}^{z}_{i};\widehat{\theta}^{LL})$, $\phi(HL;\boldsymbol{X}^{z}_{i};\widehat{\theta}^{HL})$ and $\phi(LH;\boldsymbol{X}^{z}_{i};\widehat{\theta}^{LH})$. Colors refer to $\phi(HH;\boldsymbol{X}^{z}_{i};\widehat{\theta}^{HH})$ \label{fig: indps3d}}
	\end{subfigure}
	\caption{Individual propensity score \label{fig: indps}}
\end{figure}

Given the vector of estimated parameters $\widehat{\theta}^{z}=\{\widehat{\boldsymbol{\beta}}_{HL} \cup \widehat{\boldsymbol{\beta}}_{LH} \cup \widehat{\boldsymbol{\beta}}_{HH} \}$, we denote the estimated \emph{individual propensity score} corresponding to the actual treatment $Z_{i}$ as $\widehat{\Phi_{i}}=\phi(Z_{i};\boldsymbol{X}^{z}_{i};\widehat{\theta}^{z})$. We include in $\boldsymbol{X}^{z}_{i}$ the whole set of covariates $\mathcal{X}$ we have described in Section \ref{data}.
Figures \ref{fig: indpsh} and \ref{fig: indps3d} provide a graphical intuition of the marginal and joint distribution of predicted propensity scores.
\subsubsection{Neighborhood Propensity Score \label{neighpsestim}}
In the considered empirical scenario, the neighborhood treatment is a quadrivariate continuous variable, $\boldsymbol{G}_{ct}$. 
\\

We first apply a transformation on each component of the neighborhood multi treatment (more details can be found in the Appendix \ref{norm}) so that, after the transformation, we can state that the obtained variables $G^{*}_{i,z}$ follow a normal distribution. Specifically, the four transformed components jointly follow a quadrivariate-normal distribution:
$$\boldsymbol{G}^{*}_{i} \sim \mathcal{MN}(\mathcal{\boldsymbol{\mu}}_{\boldsymbol{G}_{i}^{*}},\boldsymbol{\Sigma}_{\boldsymbol{G}^{*}}),$$
where the vector of the means $\boldsymbol{\mu}_{\boldsymbol{G}_{i}^{*}}$ depends on the individual treatment and on units' covariates through some parameters,
$$\begin{aligned}
\boldsymbol{\mu}_{\boldsymbol{G}_{i}^{*}}=&\big[\mu_{G^{*}_{i,LL}},\mu_{G^{*}_{i,B1}},\mu_{G^{*}_{i,B2}},\mu_{G^{*}_{i,C}}\big]\\
&\big[\alpha_{G^{*}_{LL}}+ \boldsymbol{\beta}^{T}_{G^{*}_{LL}}\textbf{X}^{g}_{i}+ \boldsymbol{\beta}^{T}_{G^{*}_{LL}}Z_{i},\alpha_{G^{*}_{HL}}+ \boldsymbol{\beta}^{T}_{G^{*}_{HL}}\textbf{X}^{g}_{i}+ \boldsymbol{\beta}^{T}_{G^{*}_{HL}}Z_{i},\\ & \alpha_{G^{*}_{LH}}+ \boldsymbol{\beta}^{T}_{G^{*}_{LH}}\textbf{X}^{g}_{i}+ \boldsymbol{\beta}^{T}_{G^{*}_{LH}}Z_{i},\alpha_{G^{*}_{HH}}+ \boldsymbol{\beta}^{T}_{G^{*}_{HH}}\textbf{X}^{g}_{i}+ \boldsymbol{\beta}^{T}_{G^{*}_{HH}}Z_{i}\big]
\end{aligned}
$$
and the variance-covariates matrix looks like
$$\boldsymbol{\Sigma}_{\boldsymbol{G}^{*}}=\left(\begin{array}{cccc} \sigma^{2}_{G^{*}_{LL}} & \rho_{(G^{*}_{LL},G^{*}_{HL})}   \sigma_{G^{*}_{LL}} \sigma_{G^{*}_{HL}} & \rho_{(G^{*}_{LL},G^{*}_{LH})}   \sigma_{G^{*}_{LL}} \sigma_{G^{*}_{LH}} &  \rho_{(G^{*}_{LL},G^{*}_{HH})} \sigma_{G^{*}_{LL}} \sigma_{G^{*}_{HH}}   \\ \rho_{(G^{*}_{LL},G^{*}_{HL})} \sigma_{G^{*}_{LL}} \sigma_{G^{*}_{HL}} &\sigma^{2}_{G^{*}_{HL}} & \rho_{(G^{*}_{HL},G^{*}_{LH})}   \sigma_{G^{*}_{HL}} \sigma_{G^{*}_{LH}} & \rho_{(G^{*}_{HL},G^{*}_{HH})}  \sigma_{G^{*}_{HL}} \sigma_{G^{*}_{HH}}  \\
\rho_{(G^{*}_{LL},G^{*}_{LH})} \sigma_{G^{*}_{LL}} \sigma_{G^{*}_{LH}} &\rho_{(G^{*}_{HL},G^{*}_{LH})} \sigma_{G^{*}_{HL}} \sigma_{G^{*}_{LH}}& \sigma^{2}_{G^{*}_{LH}} & \rho_{(G^{*}_{LH},G^{*}_{HH})}  \sigma_{G^{*}_{LH}} \sigma_{G^{*}_{HH}}  \\
\rho_{(G^{*}_{LL},G^{*}_{HH})}  \sigma_{G^{*}_{LL}} \sigma_{G^{*}_{HH}} & \rho_{(G^{*}_{HL},G^{*}_{HH})} \sigma_{G^{*}_{HL}} \sigma_{G^{*}_{HH}} & \rho_{(G^{*}_{LH},G^{*}_{HH})}  \sigma_{G^{*}_{LH}} \sigma_{G^{*}_{HH}}  &  \sigma^{2}_{G^{*}_{HH}} 
\\
\end{array}\right).$$
We fit \emph{Multivariate Multiple Linear Regression Model}, (\cite{davis1982distribution},\cite{duchesne2010computing}), regressing the (transformed) unit neighborhood treatment $\boldsymbol{G}^{*}_{i}$ on the individual treatment $Z_{i}$ and on the predictors that are candidate to influence the neighborhood treatment, $\boldsymbol{X}^{g}_{i}$. Here we include as explanatory variables $\boldsymbol{X}^{g}_{i}$ the whole set of characteristics $\mathcal{X}$, the individual treatment $Z_{i}$ and a measure of vertex centrality. This procedure determines $\boldsymbol{\widehat{\mu}}_{G^{*}_{i}}$. 
\\
The variance-covariance matrix is estimated looking at the residuals of the model. In particular, we first compute residuals of the model and, then, we estimate the variance and covariance matrix of the residuals $\boldsymbol{\widehat{\Sigma}}_{G^{*}}$, that results to be an unbiased estimator of $\boldsymbol{\Sigma_{G^{*}}}$. Therefore, the neighborhood propensity score corresponds to the quantity
$$\widehat{\Lambda}(\boldsymbol{g}; z,\boldsymbol{X}^{g}_{i})=\frac{1}{(2\pi)^{\frac{3}{2}}\big|\boldsymbol{\widehat{\Sigma}}_{G^{*}}\big|^{\frac{1}{2}}}\exp\bigg[-\frac{1}{2}\big(\boldsymbol{g}-\boldsymbol{\widehat{\mu}}_{G^{*}}\big)^{T}\boldsymbol{\widehat{\Sigma}}_{G^{*}}^{-1}\big(\boldsymbol{g}-\boldsymbol{\widehat{\mu}}_{G^{*}}\big)^{T}\bigg].$$

\section{Empirical Results \label{results}}
In this section, we illustrate the main empirical findings of this work \footnote{Here, we just present conclusions about the causal effects of interest, more detailed results about the models we implemented in the whole analysis can be found in the Appendix \ref{models}. Descriptives about included covariates are provided by Appendix \ref{desc}.}. We evaluate the impact of immigration policies on crime rates evaluating pairwise comparisons between the four treatment levels.  To assess the robustness of results with respect to different assumptions on the influence structure, we check the following configurations of the Influence Inputs Weights (IIW): i) $\alpha= \beta = \frac{1}{2}$, \emph{(gc)}: both geographical proximity and cultural similarity shape dependencies between units, and contribute in determining the influence index with equal weight;  ii) $\alpha=1 ,\beta = 0$, \emph{(g)}: only \emph{geographical proximity} drives interference; iii) $\alpha=0 ,\beta = 1$, \emph{(c)}: the influence structure depends on cultural similarity only and iv) $\alpha=0 ,\beta = 0 $, \emph{(noint)}: \emph{no interference} mechanism comes into play. 
\\

\begin{figure}[H]
	\centering
	\includegraphics[width=150mm,height=100mm]{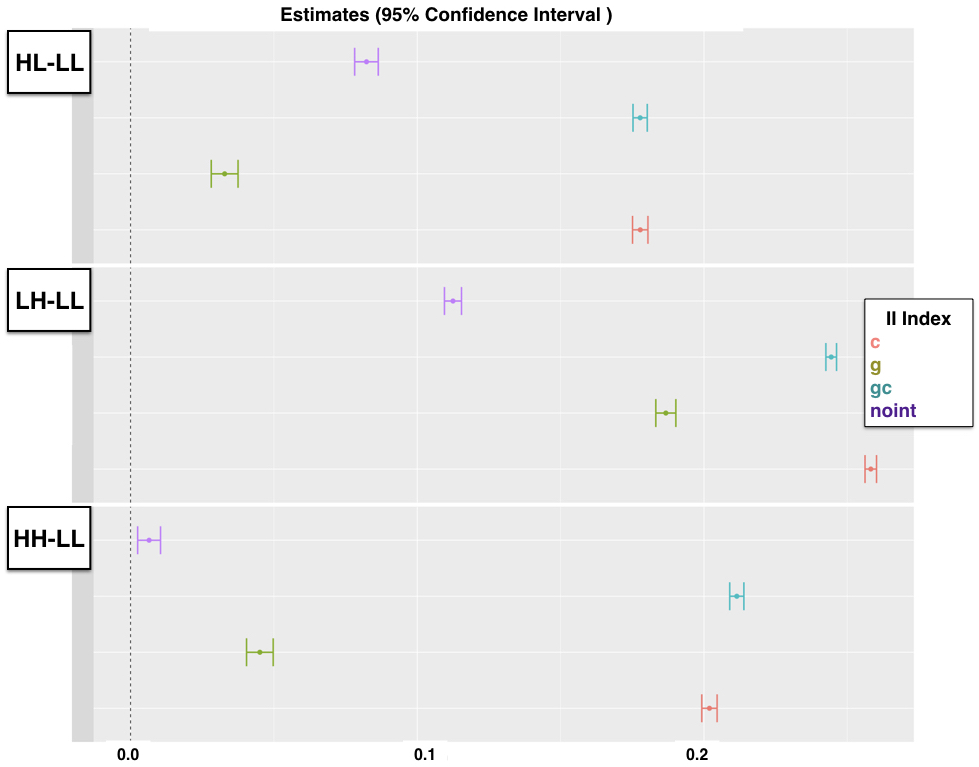}
	\caption{Direct Treatment Effects: point estimates and 95\% Confidence intervals. Colors signal the different assumption about interference: \emph{gc}(lightblue), \emph{g}(green), \emph{c}(red), \emph{noint}(purple) \label{fig: treffresults4forest}}
\end{figure}

Figures \ref{fig: treffresults4forest} graphically shows the main empirical results, which are numerically reported in Table \ref{tab: res4}. The general conclusion is that severe approaches towards immigration imply higher crime rates, compared with a welcoming political receipt. This finding holds when the comparison is with strategies with restrictive regulations only (\emph{HL-LL}), systems where only control protocols are particularly strict (\emph{LH-LL}) and profiles adopting a restrictive legislative plan in terms of both regulations and control mechanisms (\emph{HH-LL}).
If we look at how results changes with different definitions of the influence weights, we can state that ignoring the possible spillover mechanism (\emph{noint}) leads to a downward bias in the estimates. This conclusion is stable in all the contrasts of interest. On the contrary, allowing for the presence of interference increases the size of the effects. In particular, introducing the cultural similarity in the mechanism of dependencies enhances the effects' intensities (\emph{c}). Geographical proximity mitigates the impact of interference on results, but also assuming that geography is the only prompter of the spillover mechanism steers to stronger conclusions, compared to the no-interference scenario (\emph{g} and \emph{gc}). These considerations hold in all the considered comparisons. 
\begin{table}[H]
	\caption{  Direct Treatment Effects for the contrasts of interest: point estimates and 95\% Confidence intervals \label{tab: res4}}
	\scriptsize
	\renewcommand*{\arraystretch}{1.5}
	\centering
	\begin{tabular}{|l@{\hspace{0.05cm}}c@{\hspace{0.05cm}}c@{\hspace{0.05cm}}c|}
		\specialrule{2pt}{0pt}{0pt}	
		
		\multicolumn{4}{|c|}{Effects of Interest}  \\
		IIW 	& HL-LL & LH-LL & HH-LL \\ 
		$(\alpha,\beta)$ & & &  \\
		\specialrule{2pt}{0pt}{0pt}	 
		$(\frac{1}{2},\frac{1}{2}) $ &  0.17774 ***& 0.24439 ***& 0.21145 ***\\ 
		& (0.17501;0.18008) & (0.24246;0.24618) & (0.20907;0.21409)  \\ 
		
		\hline 
		$(1,0)$& 0.03281 ***& 0.1867 ***& 0.0451 ***\\
		& (0.02768;0.03721) & (0.18308;0.19007) & (0.04062;0.05006)  \\ 
		
		\hline 
		$(0,1)$& 0.17778 ***& 0.25819 ***& 0.20191 ***\\ 
		& (0.17483;0.1803) & (0.2561;0.26012) & (0.19934;0.20476)  \\ 
		
		\hline 
		$(0,0)$& 0.08228 ***& 0.11245 ***& 0.00647 ***\\ 
		& (0.07842;0.08657) & (0.10927;0.11517) & (0.00213;0.01038)  \\

		\specialrule{2pt}{0pt}{0pt}
	\end{tabular}
\end{table}
As we fully discuss in Appendix \ref{altresults}, these results are robust to different specifications of the multi-valued treatment ( we introduce an alternative definition of the multi-valued treatment collapsing the LH and HL categories into one M category ).

\section{Concluding Remarks and Discussion}

This work extends the existing framework of causal inference under interference allowing for a multi-valued treatment and for an interference structure shaped through a weighted network. This is a very common setting that can be found in a wide wide ensemble of applications. For example, political science often deals with policy evaluation settings with a multi-valued strategy, as treatments vary across multiple dimensions, so calling for an high level of complexity. Here we evaluate the effect of the national immigration policy on the crime rate. Given the multi-valued nature of the individual treatment, the neighborhood exposure cannot be summarized by a single measure, as in the binary setting. Our idea is to introduce a multi-valued network exposure, where each unit is exposed to their neighbors' treatment, weighted by the strength of their interaction. Information about the whole exposure mapping is depicted by the Neighborhood Treatments Exposure Matrix (NTEM). This framework implies an extended definition of the joint propensity score, called Joint Multiple Generalized Propensity Score (JMGPS), which models a multi-valued individual treatment and a multivariate neighborhood treatment. Direct effects of interest are pairwise comparisons of all treatment levels and they are computed comparing imputed potential outcomes controlling for the multi-valued network exposure. Our empirical results show that implementing a welcoming immigration policy causes a reduction in the crime rate. These findings suggest that welcoming immigration policies may contribute in reducing the social unrest between immigrants and natives. One possible explanation is that adopting a legislative system, which allows migrants to be actively involved in the hosting community, conceding them civil and social rights, encourages the integration process and reduces frictions. Results also show that Ignoring multi-valued interference leads to weaker estimates.

\section*{Acknowledgement}
Irene Crimaldi, Fabrizia Mealli and Costanza Tort\'u are members of the Italian Group ``Gruppo Nazionale per l’Analisi Matematica, la Probabilit\'a  e le loro Applicazioni" of the Italian Institute “Istituto Nazionale di Alta Matematica”. We are grateful for comments to participants at the ``European Cooperation for Statistics and Network Data Science" (COSTNET) workshop, at the Seventh International Workshop on Social Network Analysis (ARS '19) and at the ``Inclusive and Sustainable Development" workshop .
\section*{Declaration}
All the authors developed the methodology and the theoretical results and contributed to the final version of the manuscript.
Costanza Tortù conceived the empirical application and, under the supervision of Laura Forastiere, implemented the analysis of the data.

\section*{Funding Status}
Irene Crimaldi is partially supported by the Italian “Programma di Attività Integrata” (PAI), project “TOol for Fighting FakEs” (TOFFE) funded by IMT School for Advanced Studies Lucca.
Costanza Tortù is supported by the Frontier Proposal Fellowship (FPF) funded by IMT School for Advanced Studies Lucca.

\bibliographystyle{elsarticle-harv}
\bibliography{preprint.bib} 
\appendix
\label{app}
\section{Proofs}
\label{proofs}
\subsection{Balancing property of JMGPS}
\label{Proof 1}
We have to prove that 
$$ P(Z_{i}=z, \boldsymbol{G}_{i}= \boldsymbol{g}|\boldsymbol{X}_{i})=P(Z_{i}=z, \boldsymbol{G}_{i}= \boldsymbol{g}|\psi(z,\boldsymbol{g};\boldsymbol{X}_{i})).
$$
The expression on the leften side exactly equals JMGPS, by definition, that is $$P(Z_{i}=z, \boldsymbol{G}_{i}= \boldsymbol{g}|\boldsymbol{X}_{i})= \psi(z,\boldsymbol{g};\boldsymbol{X}_{i}).$$
We focus now on the righten side. By iterated equation we have that 
\begin{align}
P(Z_{i}=z, \boldsymbol{G}_{i}& = \boldsymbol{g}|\psi(z,\boldsymbol{g};\boldsymbol{X}_{i}))=\mathbb{E}_{\boldsymbol{X}}\big[P(Z_{i}=z, \boldsymbol{G}_{i}= \boldsymbol{g}|\boldsymbol{X}_{i},\psi(z,\boldsymbol{g};\boldsymbol{X}_{i}))|\psi(z,\boldsymbol{g};\boldsymbol{X}_{i})\big] \nonumber\\
& =\mathbb{E}_{\boldsymbol{X}}\big[P(Z_{i}=z, \boldsymbol{G}_{i}= \boldsymbol{g}|\boldsymbol{X}_{i})|\psi(z,\boldsymbol{g};\boldsymbol{X}_{i})\big] \nonumber\\
& =\mathbb{E}_{\boldsymbol{X}}\big[\psi(z,\boldsymbol{g};\boldsymbol{X}_{i})|\psi(z,\boldsymbol{g};\boldsymbol{X}_{i})\big]=\psi(z,\boldsymbol{g};\boldsymbol{X}_{i}) \nonumber
\end{align} 
The second equality holds as the joint multiple generalized propensity score, by definition, is functionally related to the characteristics $\boldsymbol{X}_{i}$. The third equality follows from the definition of JMGPS.  
\\

Both above expressions are equal to the joint multiple propensity score itself and, hence, they are also equal to each other.

\subsection{Conditional unconfoundness of $D_{i}(z)$ and $G_{i}$ given JMGPS}
\label{Proof 2}
We have to show that
$$P(D_{i}(z)=1, \boldsymbol{G}_{i}= \boldsymbol{g}|Y_{i}(z,\boldsymbol{g}),\psi(z,\boldsymbol{g};\boldsymbol{X}_{i}))=P(D_{i}(z)=1, \boldsymbol{G}_{i}= \boldsymbol{g}|\psi(z,\boldsymbol{g};\boldsymbol{X}_{i})).
$$
\textbf{Righten side}.  We first focus on the expression that lies at the righten side. By the fact that $(D_i(z)=1)=(Z_i=z)$ and  Proposition \ref{prop:bal}, we have 
$$P(D_{i}(z)=1, \boldsymbol{G}_{i}= \boldsymbol{g}|\psi(z,\boldsymbol{g};\boldsymbol{X}_{i}))=P(Z_{i}=z, \boldsymbol{G}_{i}= \boldsymbol{g}|\psi(z,\boldsymbol{g};\boldsymbol{X}_{i}))=\psi(z,\boldsymbol{g};\boldsymbol{X}_{i}).$$
\textbf{Leften side}. By iterated equations, we have  
\begin{align}
&P(D_{i}(z)=1, \boldsymbol{G}_{i}= \boldsymbol{g}|Y_{i}(z,\boldsymbol{g}),\psi(z,\boldsymbol{g};\boldsymbol{X}_{i}))
\nonumber \\
&=\mathbb{E}_{\boldsymbol{X}}\big[P\big(D_{i}(z)=1, \boldsymbol{G}_{i}= \boldsymbol{g}|\boldsymbol{X}_{i},\psi(z,\boldsymbol{g};\boldsymbol{X}_{i}),Y_{i}(z,\boldsymbol{g})\big)|Y_{i}(z,\boldsymbol{g}),\psi(z,\boldsymbol{g};\boldsymbol{X}_{i})\big]  
\nonumber\\
&= \mathbb{E}_{\boldsymbol{X}}\big[P\big(D_{i}(z)=1, \boldsymbol{G}_{i}= \boldsymbol{g}|Y_{i}(z,\boldsymbol{g}),\boldsymbol{X}_{i}\big)|Y_{i}(z,\boldsymbol{g}),\psi(z,\boldsymbol{g};\boldsymbol{X}_{i})\big]  \nonumber\\
&= \mathbb{E}_{\boldsymbol{X}}\big[P\big(D_{i}(z)=1, \boldsymbol{G}_{i}= \boldsymbol{g}|\boldsymbol{X}_{i} \big)| Y_{i}(z,\boldsymbol{g}),\psi(z,\boldsymbol{g};\boldsymbol{X}_{i})] \nonumber\\
& =\mathbb{E}_{\boldsymbol{X}}\big[\psi(z,\boldsymbol{g};\boldsymbol{X}_{i})| Y_{i}(z,\boldsymbol{g}),\psi(z,\boldsymbol{g};\boldsymbol{X}_{i})]= \psi(z,\boldsymbol{g};\boldsymbol{X}_{i}), \nonumber
\end{align} 
where the second equality is obtained taking into account that the joint multiple generalized propensity score is a function of covariates, the third equality results from applying the Assumption \ref{ass: weekunc}, while the forth equality holds recalling that $(D_{i}(z)=1)=(Z_{i}=z)$ and Definition \ref{def:JMGPS}.

\subsection{Conditional unconfoundness of $D_{i}(z)$ and $G_{i}$ given individual and neighborhood propensity scores}
\label{Proof 3}
We have to show that
$$P(D_{i}(z)=1, \boldsymbol{G}_{i}= \boldsymbol{g}|Y_{i}(z,\boldsymbol{g}),\phi(z; \boldsymbol{X}^{z}_{i}), \lambda(\boldsymbol{g}; z,\boldsymbol{X}^{g}_{i}))=P(D_{i}(z)=1, \boldsymbol{G}_{i}= \boldsymbol{g}|\phi(z; \boldsymbol{X}^{z}_{i}), \lambda(\boldsymbol{g}; z,\boldsymbol{X}^{g}_{i})),
$$
where $\phi(z ;\boldsymbol{X}^{z}_{i})=P(D_{i}(z)=1|\boldsymbol{X}^{z}_{i} )$ and
$\lambda(\boldsymbol{g}; z,\boldsymbol{X}^{g}_{i})=P(\boldsymbol{G}_{i}=\boldsymbol{g}|Z_i=z,\boldsymbol{X}^{g}_{i})$. We proceed showing that both sides of the equation are equal to the joint multiple generalized propensity score.
\\

\textbf{Righten side}. By iterated equations, we have
\begin{align}
&P(D_{i}(z)=1, \boldsymbol{G}_{i}= \boldsymbol{g}|\phi(z; \boldsymbol{X}^{z}_{i}), \lambda(\boldsymbol{g}; z,\boldsymbol{X}^{g}_{i})) \nonumber\\  
&=\mathbb{E}_{\boldsymbol{X}}\big[P(D_{i}(z)=1, \boldsymbol{G}_{i}=\boldsymbol{g}|\boldsymbol{X}_{i}, \phi(z ;\boldsymbol{X}^{z}_{i}), \lambda(\boldsymbol{g}; z,\boldsymbol{X}^{g}_{i})) | \phi(z ;\boldsymbol{X}^{z}_{i}), \lambda(\boldsymbol{g}; z,\boldsymbol{X}^{g}_{i}) \big]  \nonumber \\
&=\mathbb{E}_{\boldsymbol{X}}\big[P(D_{i}(z)=1, \boldsymbol{G}_{i}=\boldsymbol{g}|\boldsymbol{X}_{i}) |\phi(z; \boldsymbol{X}^{z}_{i}), \lambda(\boldsymbol{g}; z,\boldsymbol{X}^{g}_{i})\big]   \nonumber \\
&=\mathbb{E}_{\boldsymbol{X}}\big[ \psi(z,\boldsymbol{g};\boldsymbol{X}_{i})|\phi(z; \boldsymbol{X}^{z}_{i}), \lambda(\boldsymbol{g}; z,\boldsymbol{X}^{g}_{i})\big] = \psi(z,\boldsymbol{g};\boldsymbol{X}_{i}). \nonumber 
\nonumber
\end{align} 
The above equalities result from the fact that both $\phi(z; \boldsymbol{X}^{z}_{i})$ and $\lambda(\boldsymbol{g}; z,\boldsymbol{X}^{g}_{i})$ are function of $\boldsymbol{X}_{i}$ (second equality) and from the factorization $\psi(z,\boldsymbol{g};\boldsymbol{X}_{i})=\phi(z; \boldsymbol{X}^{z}_{i}) \lambda(\boldsymbol{g}; z,\boldsymbol{X}^{g}_{i})$ (third equality).
\\

\textbf{Leften side}. By iterated equations, we have
\begin{align}
&P(D_{i}(z)=1, \boldsymbol{G}_{i}= \boldsymbol{g}|Y_{i}(z,\boldsymbol{g}),\phi(z; \boldsymbol{X}^{z}_{i}), \lambda(\boldsymbol{g}; z,\boldsymbol{X}^{g}_{i})) \nonumber\\  
&=\mathbb{E}_{\boldsymbol{X}}\big[P(D_{i}(z)=1, \boldsymbol{G}_{i}=\boldsymbol{g}|\boldsymbol{X}_{i},Y_{i}(z,\boldsymbol{g}),\phi(z ;\boldsymbol{X}^{z}_{i}), \lambda(\boldsymbol{g}; z,\boldsymbol{X}^{g}_{i})) | Y_{i}(z,\boldsymbol{g}),\phi(z ;\boldsymbol{X}^{z}_{i}), \lambda(\boldsymbol{g}; z,\boldsymbol{X}^{g}_{i}) \big]  \nonumber \\
&=\mathbb{E}_{\boldsymbol{X}}\big[P(D_{i}(z)=1, \boldsymbol{G}_{i}=\boldsymbol{g}|\boldsymbol{X}_{i},Y_{i}(z,\boldsymbol{g})| Y_{i}(z,\boldsymbol{g}),\phi(z ;\boldsymbol{X}^{z}_{i}), \lambda(\boldsymbol{g}; z,\boldsymbol{X}^{g}_{i}) \big]  \nonumber \\
&=\mathbb{E}_{\boldsymbol{X}}\big[P(D_{i}(z)=1, \boldsymbol{G}_{i}=\boldsymbol{g}|\boldsymbol{X}_{i}) |Y_{i}(z,\boldsymbol{g}),\phi(z; \boldsymbol{X}^{z}_{i}), \lambda(\boldsymbol{g}; z,\boldsymbol{X}^{g}_{i})\big]   \nonumber \\
&=\mathbb{E}_{\boldsymbol{X}}\big[ \psi(z,\boldsymbol{g};\boldsymbol{X}_{i}) |Y_{i}(z,\boldsymbol{g}),\phi(z; \boldsymbol{X}^{z}_{i}), \lambda(\boldsymbol{g}; z,\boldsymbol{X}^{g}_{i})\big] \nonumber \\
& = \psi(z,\boldsymbol{g};\boldsymbol{X}_{i}), \nonumber
\end{align} 
where the second equality results from the fact that the two propensity scores are function of the covariates, the third equality comes from Assumption \ref{ass: weekunc}, the fourth equality is obtained recalling Definition \ref{def:JMGPS} and that $(D_{i}(z)=1)=(Z_{i}=z)$ and, finally, the last equality follows from the factorization of the JMGPS.

\section{Influence Index detailed construction \label{iciapp}}
The Influence index (I) is formally defined as 
$$I_{cc',t}=\alpha \times IG_{cc'}+\beta \times IC_{cc',t}$$
where $IG$ denotes the geographical proximity indicator while $IC$ states for the cultural similarity indicator. The former is time invariant, while the latter provides a temporal variation. Here, we discuss the detailed construction of these two indexes, which determine the interference structure. 
\\

\begin{figure}[H]
	\begin{subfigure}{.5\textwidth}
		\centering
		\includegraphics[width=85mm]{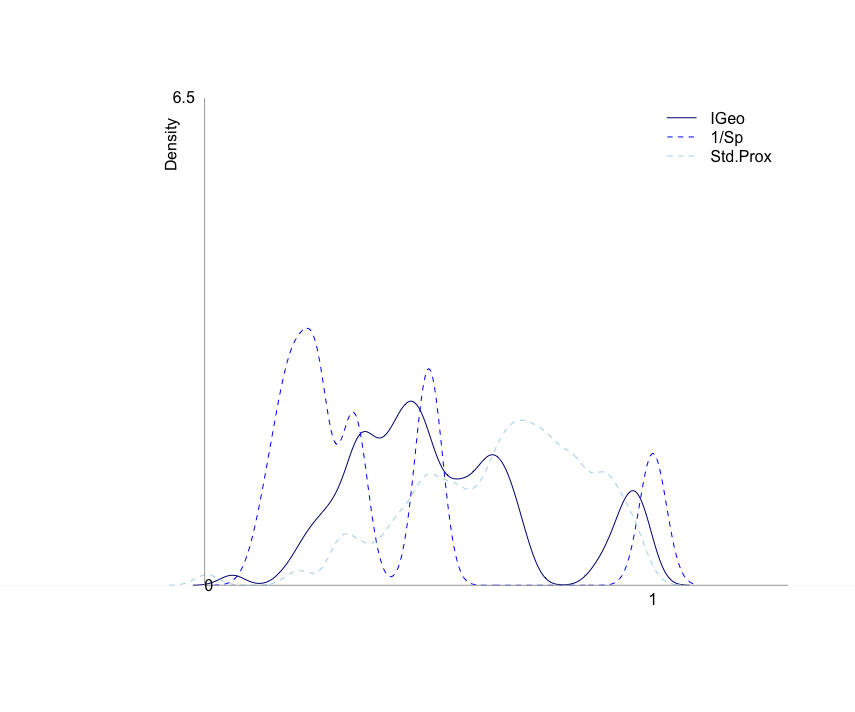}
		\caption{Geographical Proximity Indicator \label{fig: icicomp_geo}}
	\end{subfigure}
	\begin{subfigure}{.5\textwidth}
		\centering
		\includegraphics[width=85mm]{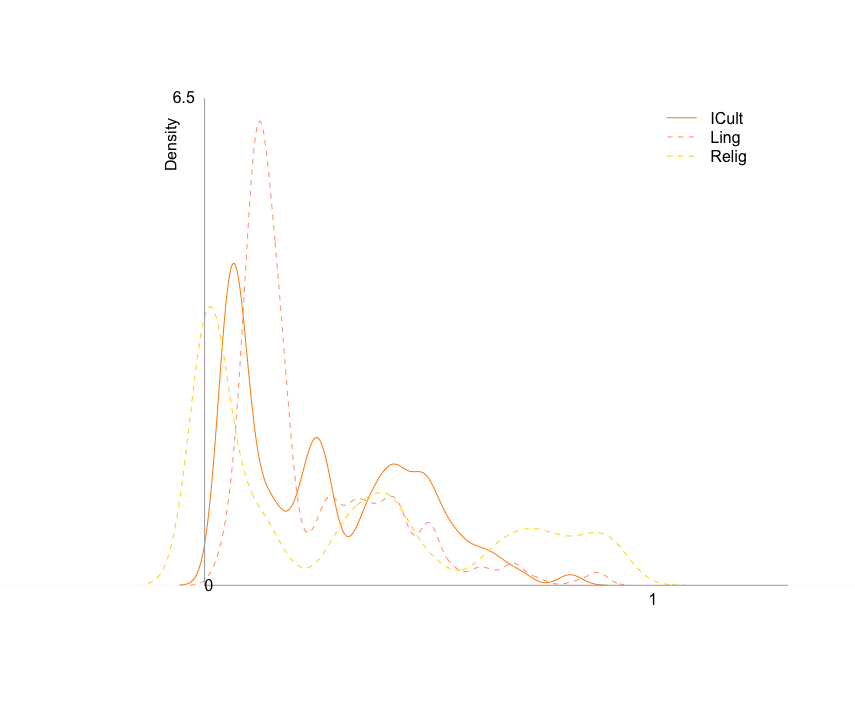}
		\caption{Cultural similarity Indicator \label{fig: icicomp_cult}}
	\end{subfigure}
	\caption{Influence index components density distributions  \label{fig: icicomp}}
\end{figure}

We build up the geographical proximity index taking into account of two variables: a boundaries-related variable and a geographical distance-related variable. The former, that we denote by $Sp$ counts the minimum number of states one needs to cross by, at the aim of reaching country $c'$ starting from country $c$. Thus, if we consider a graph collecting all the national states, this variable represents the length of the shortest path between $c$ and $c'$ \footnote{We assume that the pairs of countries France and Great Britain, Ireland and Great Britain share a common boundary, as, even if they're formally separated by the English Channel and the Irish Sea, respectively, they are very near and connections are extremely simple}. The latter, that we denote as $Dist^{std}$ is a standardized measure of geographic distance between the most populated cities belonging to the two countries. Formally, the geographical proximity indicator is computed as follows
$$IG_{cc'}=0.5 \times \frac{1}{\text{Sp}_{cc'}} + 0.5 \times (1-\text{Dist}^{std}_{cc'})=0.5 \times \frac{1}{\text{Sp}_{cc'}} + 0.5 \times \text{Prox}^{std}_{cc'}$$
On the other side, the cultural similarity indicator measures the level of cultural similarity between two countries $c$ and $c'$ at a given time $t$. We summarize this aspect evaluating the linguistic similarity and the religious similarity, through the variables $Ling$ and $Relig$. These measures have been defined by the CEPII Linguistic Dataset \cite{ melitz2014native} and CEPII Gravidata Dataset (\cite{fouquin2016two}), respectively. The \emph{linguistic proximity indicator} gives a unique measure of how much the whole linguistic systems differ in the two countries, both in terms of the distribution of spoken languages over the population and in terms of the linguistic roots. The \emph{religious similarity indicator} takes into account of the distribution of practised religions: an high value of this variable signals an high similarity in terms of prevalence of the various religious communities at time $t$.
$$IC_{cc',t}=0.5 \times \text{Ling}_{cc',t} + 0.5 \times \text{Relig}_{cc',t}$$
Figure \ref{fig: icicomp} shows the density distributions of the two indicators that contribute in determining the Influence Index, as well as of their respective sub-components.

\section{Transformation of the NTEM components \label{norm}}
\begin{figure}[H]
	\centering
	\begin{subfigure}{.49\textwidth}
		\includegraphics[width=70mm,height=80mm]{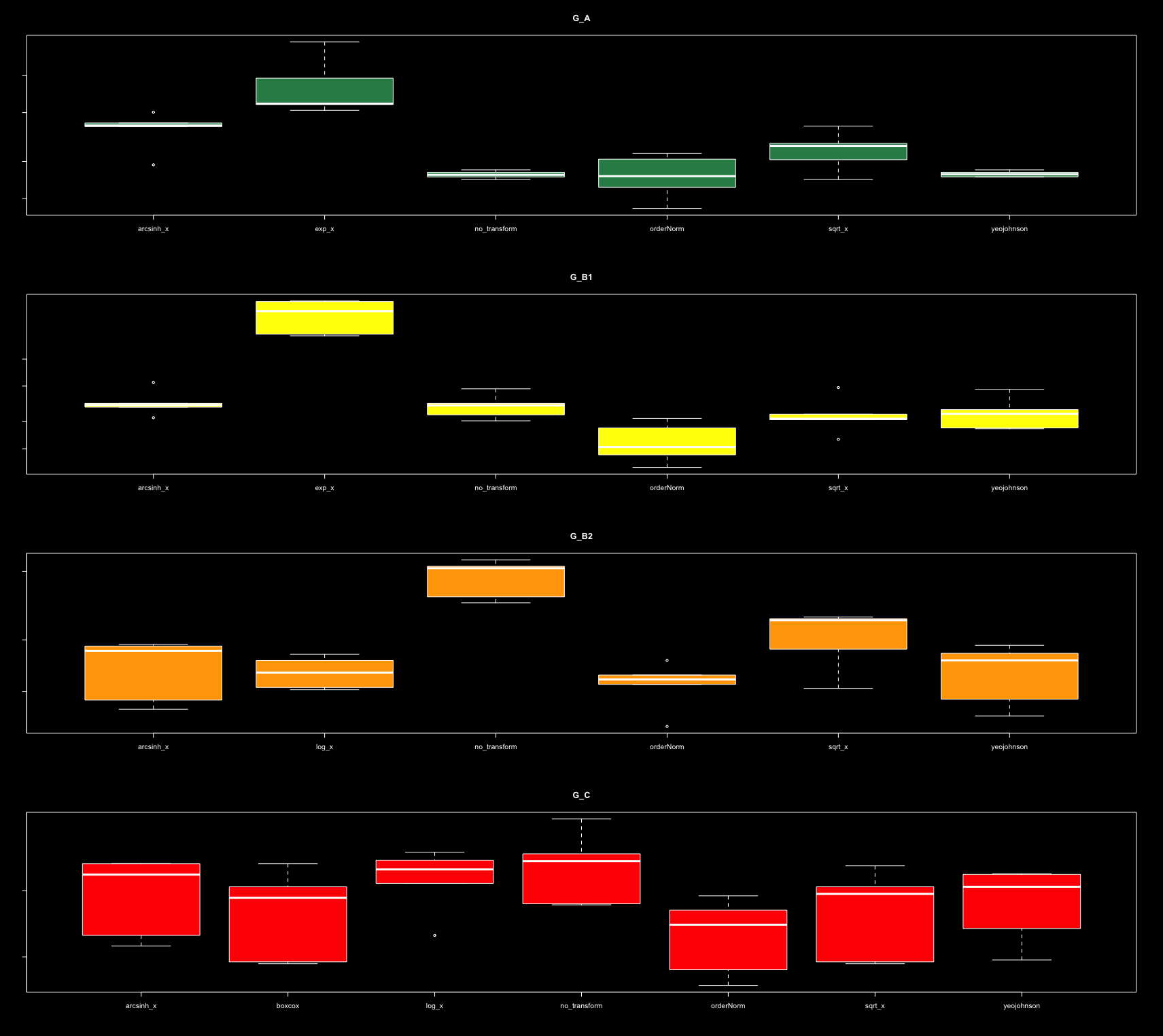}
		\caption{Best Normalizing methods: comparison \label{fig: bn_methcompare}}
	\end{subfigure}\hspace{0.1cm}
	\begin{subfigure}{.49\textwidth}
		\includegraphics[width=83mm,height=79mm]{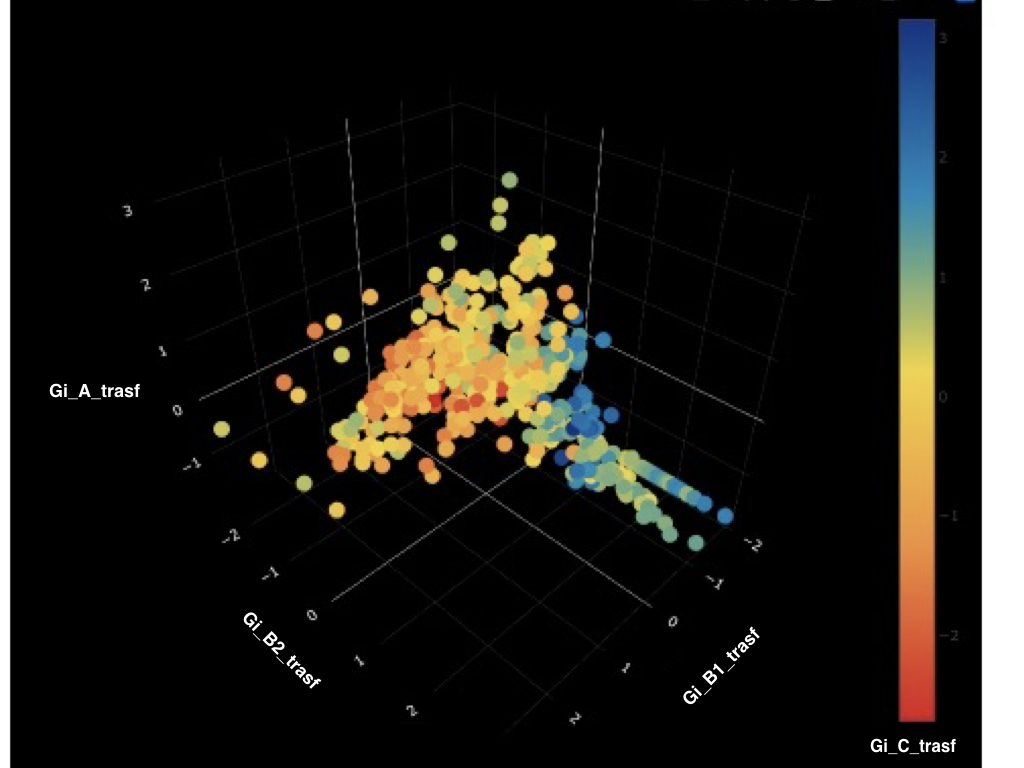}
		\caption{Tridimensional plot of trasformed  variables \label{fig: bn_3dgif_trasf}}
	\end{subfigure}
	\caption{Best Normalizing method \label{fig: bn}}
\end{figure}

We run some checks about the normality of the components $G_{i,z}$. The Shapiro Tests for Normality (\cite{shapiro1965analysis}), separately  conducted on the four components, $G_{i,LL}$, $G_{i,HL}$, $G_{i,LH}$, $G_{i,HH}$, rejects the Normality null-Hypothesis.
\\
Hence, we decide to apply a transformation to each of the $G_{i,z}$. We conduct some tests experimenting various transformation methods and we compare them selecting the best approach according to the Pearson P statistic for Normality (divided by its degrees of freedom). We use repeated cross validation to estimate the out-of-sample performance of all these methods. Figure \ref{fig: bn_methcompare} shows the boxplots of the out of sample estimated normality statistics for all the techniques that we experiment, over the four variables of interest (under $\alpha=\frac{1}{2}$ and $\beta=\frac{1}{2}$).  We find out that the method that performed better in handling the $G_{i,z}$ variables is the \emph{Ordered Quantile (ORQ) transformation}, for all the various configurations of the II input weights. Figure \ref{fig: bn_3dgif_trasf} represents the tridimensional scatterplot of trasformed variables. 

Ordered Quantile trasformation  (\cite{bartlett1947use},\cite{van1952order})is based on ranks. Essentially, the values of a variable, judged as a vector, are mapped to their percentile, and then to the same percentile of the Standard Normal Distribution. As long as the number of ties is negligible, this method guarantees that the transformed variable follows a Normal Distribution. 
Formally, each variable $G_{i,z}$ is transformed according to the following formula:
$$ G^{*}_{i,z}=\Phi^{-1}\Bigg(\frac{\text{rank}(G_{iz})}{N+1}\Bigg),$$
where $\Phi$ is the cumulative density function of a Standard Normal distribution, $N$ is the number of observations. We denote as $ G^{*}_{i,z}$ the variable resulting from the  Ordered Quantile transformation.

\section{Descriptives \label{desc}}
\begin{figure}[H]
	\centering
	\includegraphics[width=100mm]{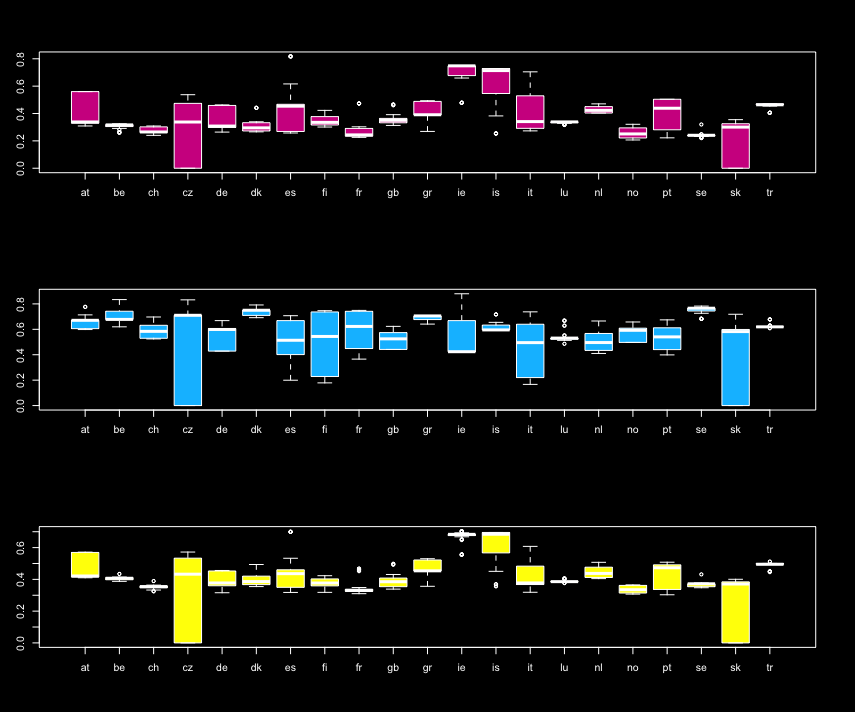}
	\caption{\emph{Reg} (violet), \emph{Control} (blue) and \emph{ImPol} (yellow) indicator, across countries \label{fig: indexescountries}}
\end{figure}

\begin{figure}[H]
	\begin{subfigure}{.5\textwidth}
		\centering
		\includegraphics[width=82mm]{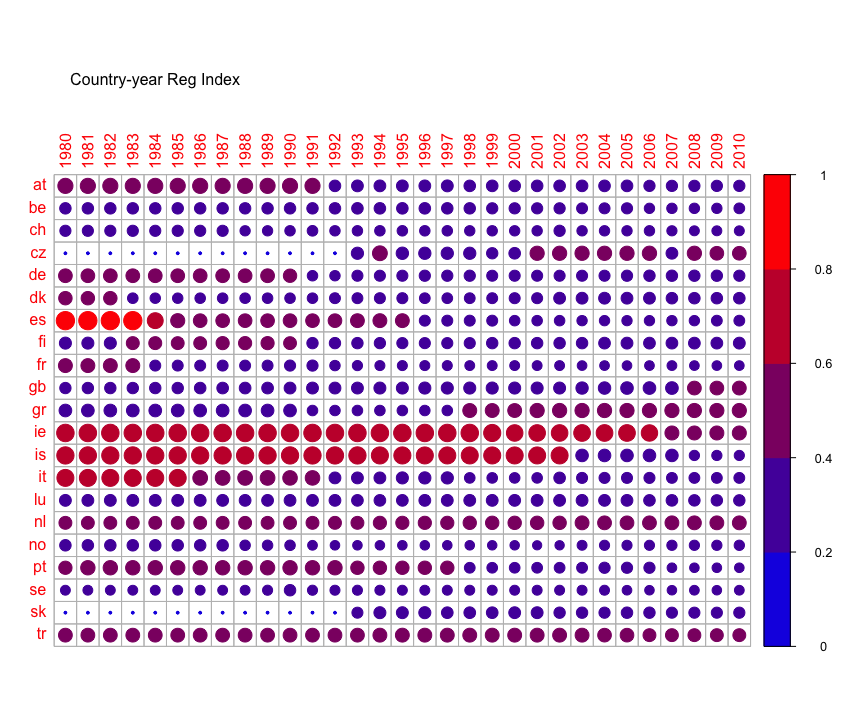}
		\caption{Regulation indicator  \label{fig: corcntryyear_reg}}
	\end{subfigure}
	\begin{subfigure}{.5\textwidth}
		\centering
		\includegraphics[width=82mm]{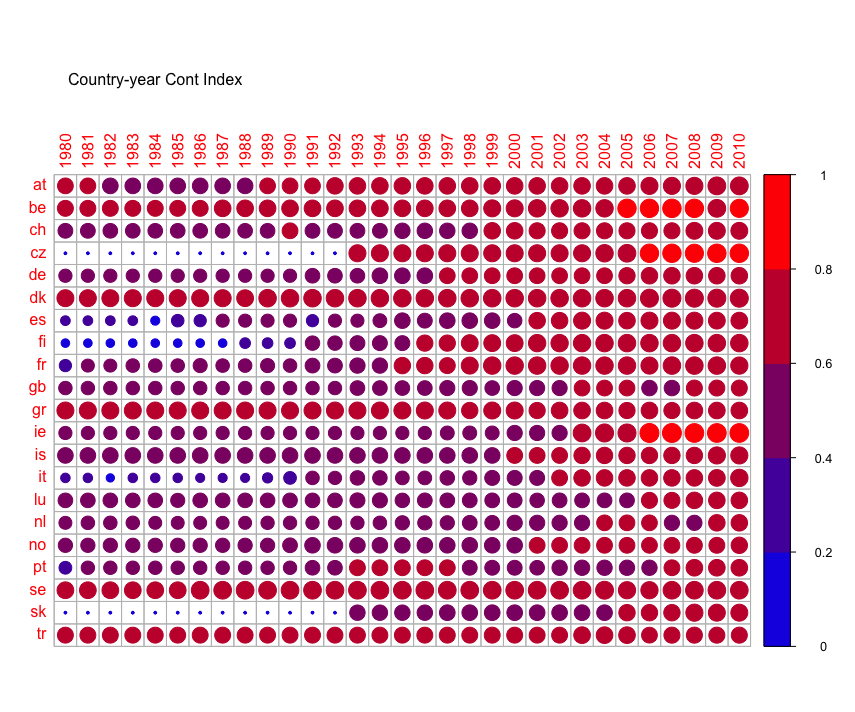}
		\caption{Control indicator  \label{fig: corcntryyear_cont}}
	\end{subfigure}
	\caption{Indicators of interest mapped over country-year profiles \label{fig: corcntryyear}}
\end{figure}

This paragraph provides some descriptives. Figure \ref{fig:interferencecomp} shows the density distributions of the indicators measuring the restrictiveness of \emph{regulations} (\emph{Reg}) and \emph{control strategies} (\emph{Cont}), over years. Regulations have become more welcoming over time while control strategies have turned to a more severe attitude.
\begin{figure}[H]
	\begin{subfigure}{.5\textwidth}
		\centering
		\includegraphics[width=85mm]{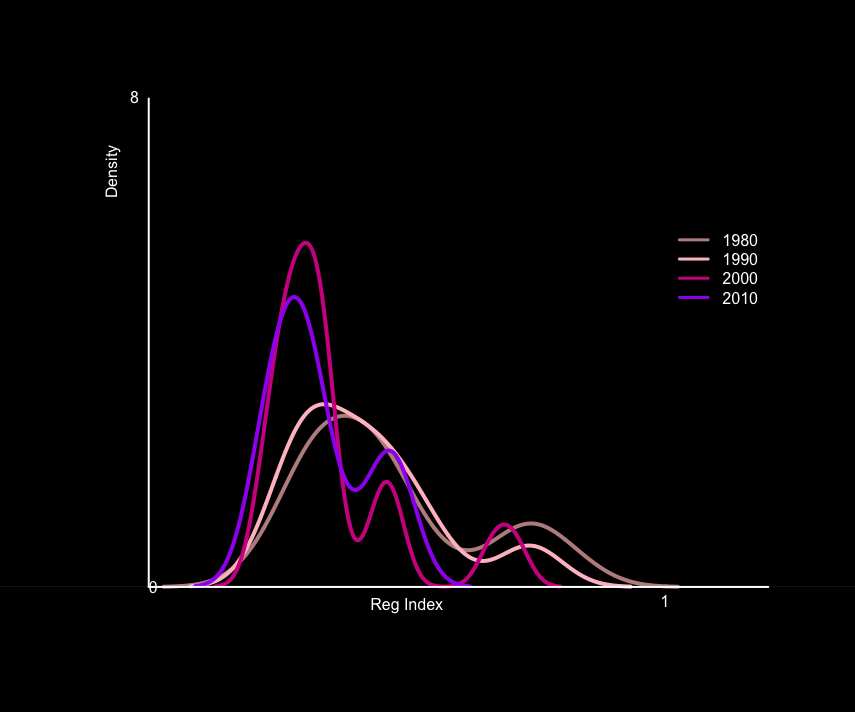}
		\caption{Regulations indicator \label{fig: indexyears_reg}}
	\end{subfigure}
	\begin{subfigure}{.5\textwidth}
		\centering
		\includegraphics[width=85mm]{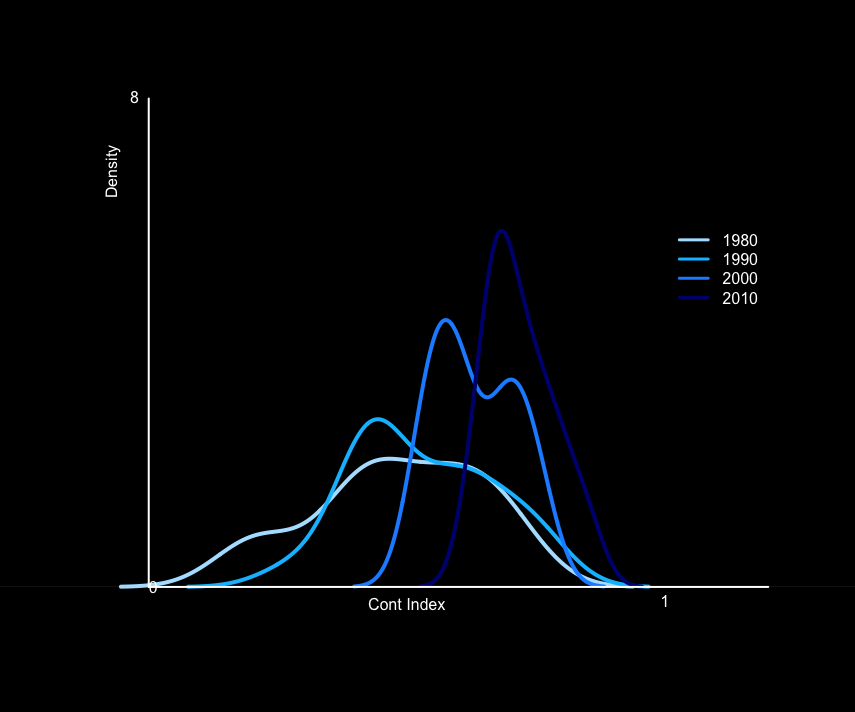}
		\caption{Control indicator  \label{fig: indexyears_cont}} 
	\end{subfigure}
	\caption{Indicators measuring the restrictiveness of immigration policies over years \label{fig: indexyears}}
\end{figure}
Figure \ref{fig: indexescountries} represents the variation of the distributions of the \emph{Reg} (violet), \emph{Control} (blue) and \emph{ImPol} (yellow) indicators in the 22 countries that we have included in the analysis.
Figure \ref{fig: corcntryyear} consents to inspect the strictness of regulations and control implemented policies in each country-year profile.
Table \ref{tab: desc} shows the basic descriptives of all the variables we have included in the analysis.
\begin{table}[H] 
	\centering 
	\scriptsize
	\renewcommand*{\arraystretch}{1.5}
	\caption{Descriptive statistics \label{tab: desc}} 
	\begin{tabular}{@{\extracolsep{5pt}} llccccccc} 
		\\[-1.8ex]\hline 
		\hline \\[-1.8ex] 
		Variable&	Var. Label & N & Mean & St. dev & Min & Pctl(25) &  Pctl(75) & Max \\ 
		\hline \\[-1.8ex] 
		Crime rate (every 10.000 inhab.)	& \textit{rate }&  612 & 1.326 & 0.636 & 0.000 & 0.910 & 1.560 & 3.430  \\ 
		Fertility rate & \textit{ferrate} &612 & 1.675 & 0.498 & 0.000 & 1.440 & 1.840 & 4.360 \\ 
		Power distributed to gender Index &	\textit{powgend} & 612 & 1.876 & 0.973 & $-$0.854 & 1.408 & 2.394 & 3.714  \\ 
		Health equality Index &	\textit{eq\_health} &  612 & 2.418 & 0.588 & 0.612 & 1.972 & 2.775 & 3.792 \\ 
		Educational inequality Gini index (/60) &	\textit{ineq\_educ} & 612 & 0.269 & 0.163 & 0.000 & 0.149 & 0.361 & 0.893 \\ 	
		Income inequality Gini index (/60)& i\textit{neq\_inc} & 612 & 0.485 & 0.139 & 0.000 & 0.432 & 0.564 & 0.867 \\ 
		Equal access index &	\textit{eq\_access} & 612 & 0.871 & 0.137 & 0.290 & 0.856 & 0.945 & 0.986  \\ 
		Equal distribution of resources index	& \textit{eq\_resdist }& 612 & 0.924 & 0.070 & 0.588 & 0.908 & 0.964 & 0.986  \\ 
		Civil partecipation index &	\textit{civilpart }& 612 & 0.641 & 0.111 & 0.161 & 0.616 & 0.690 & 0.885 \\ 
		Access to justice index	& \textit{accjust} &612 & 0.940 & 0.121 & 0.165 & 0.944 & 0.989 & 0.995 \\ 
		State ownership of economy index &	\textit{econcont} &612 & 1.219 & 0.616 & $-$0.536 & 0.890 & 1.636 & 2.731\\ 
		Freedom of expression index &	\textit{freeexp} & 612 & 0.934 & 0.128 & 0.128 & 0.955 & 0.979 & 0.991 \\ 
		Freedom of religion index & \textit{freerelig }& 612 & 1.958 & 0.751 & $-$1.003 & 1.749 & 2.519 & 2.766 \\ 
		Life expectancy (/ 100) &	\textit{lifeexp} & 612 & 0.746 & 0.126 & 0.000 & 0.749 & 0.785 & 0.824\\ 
		GDP per capita (/ 10.000)&	\textit{gdppc} & 612 & 2.666 & 1.154 & 0.610 & 1.954 & 3.373 & 8.192 \\ 
		\hline \\[-1.8ex] 
	\end{tabular} 
\end{table}

\begin{figure}[H]
	\centering
	\includegraphics[width=100mm]{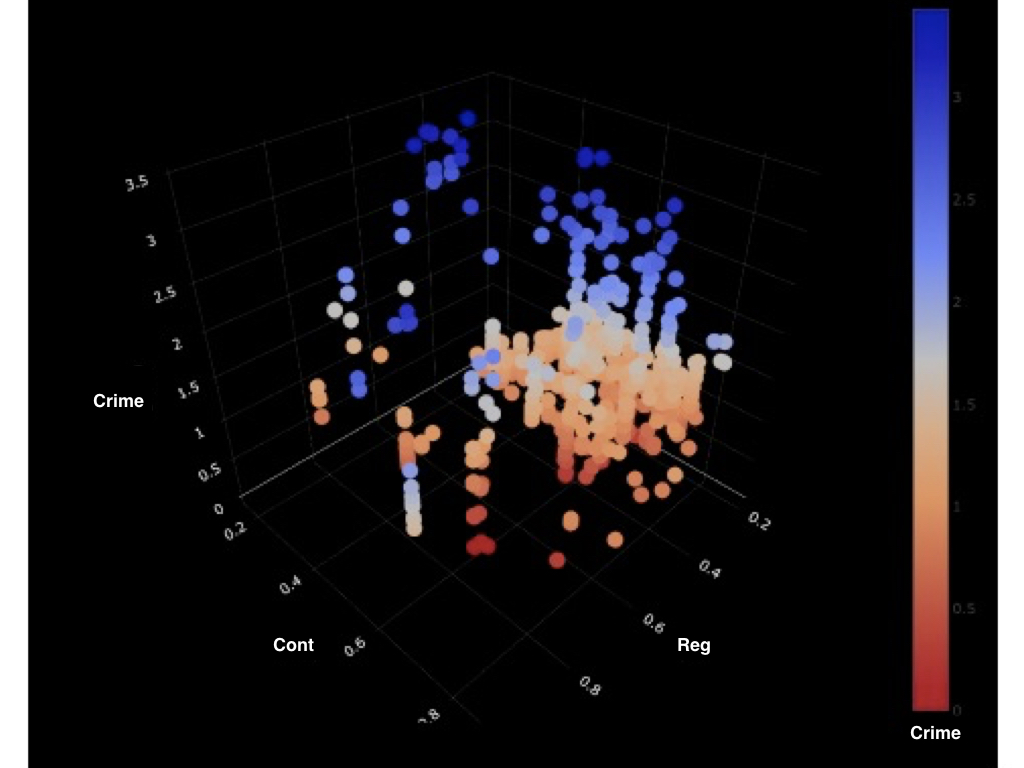}
	\caption{Tridimensional plot of the two indicators \emph{Reg} and \emph{Cont}, with respect to the \emph{Crime Rate}. 
		\label{fig: regcontcrine}}
\end{figure}

Figure \ref{fig: regcontcrine} shows the tridimensional plot of the two indicators measuring the restrictiveness towards migrants, and the corresponding the crime rate.
\section{Results under different configurations of the treatment  \label{altresults}}
\begin{figure}[H]
	\centering
	\begin{subfigure}{.49\textwidth}
		\includegraphics[width=85mm]{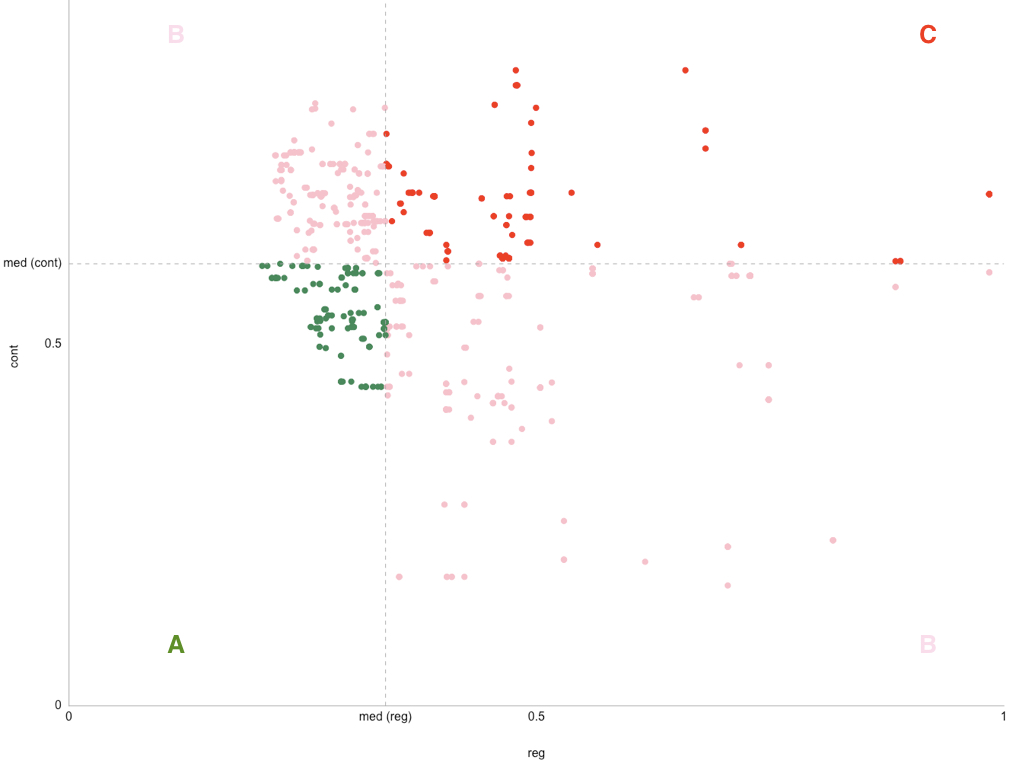}
		\caption{Multi-valued with three categories,  $Z^{(3)}_{i}$ \label{fig: trtr}}
	\end{subfigure}\hspace{0.1cm}
	\begin{subfigure}{.49\textwidth}
		\includegraphics[width=75mm]{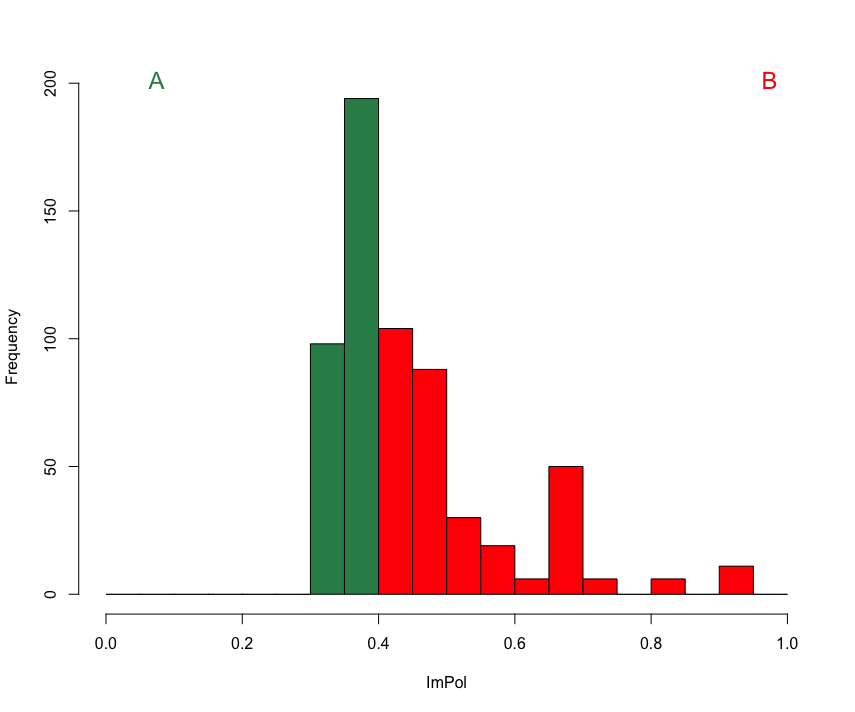}
		\caption{Binary with two categories, $Z^{(2)}_{i}$ \label{fig: bintr}}
	\end{subfigure}
	\caption{Alternative definitions of the treatment variable \label{fig: alttr}}
\end{figure}

This section shows results under alternative specifications of the treatment variable of interest. In particular, as Definition \ref{def: ind_tr32} clarifies, we test two secondary ways of detecting the treatment categories. 
\begin{Definition}{Alternative specifications of the treatment variable \label{def: ind_tr32}}
	Let us indicate $Z^{K}_{i}$ a generic treatment variable defined over $K$ categories. We consider the following treatment classifications
	\begin{enumerate}
		\item Multi-valued treatment with three categories,  $Z^{(3)}_{i}$, which have been defined collapsing the categories HL and LH of the original individual treatment variable.
		\begin{itemize}
			\item $Z^{3}_{i}$=L if $\text{reg}_{i} \le \text{med}_{reg}$ and $\text{cont}_{i} \le \text{med}_{cont}$: this category identifies profiles that are barely restrictive with respect to the two mechanisms. 
			\item $Z^{3}_{i}$=H if $\text{reg}_{i} \ge \text{med}_{reg}$ and $\text{cont}_{i} \ge \text{med}_{cont}$: this category denotes an highly restrictive policy towards migrants with respect to both regulations and control.
			\item $Z^{3}_{i}$=M otherwise \footnote{note that the A and C categories exactly coincide with the A and C categories of the four-valued treatment}
		\end{itemize}
		\item Binary treatment with two categories, $Z^{(2)}_{i}$ defined as follows
		\begin{itemize}
			\item $Z^{2}_{i}$=L if $\text{impol}_{i} \le \text{med}_{ImPol}$ 
			\item $Z^{2}_{i}$=H if $\text{impol}_{i} > \text{med}_{ImPol}$
		\end{itemize}
	\end{enumerate}
\end{Definition}
Figure \ref{fig: alttr} graphically represents these two alternative treatment characterizations. 
Table \ref{tab: altres} shows results under these two definitions of the treatment variable. As it is immediate to observe, these results are robust with the main findings of this paper.
\begin{table}[H]
	\scriptsize
	\centering
	\renewcommand*{\arraystretch}{1.5}
	\caption{Results under different treatment definitions \label{tab: altres}}
	\begin{tabular}{|l|c@{\hspace{0.05cm}}c| c@{\hspace{0.05cm}} |}
		\specialrule{2pt}{0pt}{0pt}	
		& \multicolumn{3}{c|}{Treatment categories} \\
		\hline
		& \multicolumn{2}{c|}{3 (L,M,H)} & 2 (L,H) \\
		& \multicolumn{2}{c|}{Effects of Interest} & Effects of Interest \\
		
		IIW 	& M-L & H-L & H-L\\ 
		
		$(\alpha,\beta)$ & & & \\
		\specialrule{2pt}{0pt}{0pt}	
		$(\frac{1}{2},\frac{1}{2}) $  & 0.06648 ***& 0.04986  ***& 0.03875***  \\ 
		&(0.06485;0.06815) & (0.04774;0.05196) &  (0.01424;0.06133)\\ 
		\hline
		$(1,0)$ &  0.04363  ***& 0.01781  *** & 0.04126 ***\\ 
		&(0.04203;0.04527) & (0.01573;0.01987) &  (0.01686;0.06374) \\ 
		\hline
		$(0,1)$ &0.09282  ***& 0.03523  ***&  0.03587 ***\\ 
		& (0.09228;0.09338) & (0.03452;0.03592)  & (0.01288;0.05705) \\ 
		\hline
		$(0,0)$ & 0.0727  ***& '0.0008 &  0.03506*** \\ 
		& (0.07027;0.07524) & (-0.00386;0.00303)  & (0.01443;0.05789)\\ 
		
		\specialrule{2pt}{0pt}{0pt}
	\end{tabular}
\end{table}
Taking into consideration of a binary treatment ( which we obtain simply differentiating the country-year profiles whose observed value of the general immigration policies indicator is above its reference median ) still leads to positive results, regardless of the assumption about interference. But in the last two scenarios, effects are significantly weaker.
\section{Models Results \label{models}}
\subsection{Model for Z \label{zmodel}}
\begin{table}[H] \centering 
	\scriptsize
	\renewcommand*{\arraystretch}{1.5}
	\caption{Model for the individual treatment $Z_{i}$: multinomial logit }
	\begin{tabular}{l@{\extracolsep{10pt}}l@{\extracolsep{10pt}}l@{\extracolsep{10pt}}l@{\extracolsep{10pt}} } 
		\specialrule{2pt}{0pt}{0pt}
		& \multicolumn{3}{c}{\textit{Dependent variable:}} \\ 
		\cline{2-4} 
		& \multicolumn{3}{c}{$Z_{i}$} \\ 	
		& HL & LH & HH\\
		\hline
		(Intercept) & 32.79243***(12.78648) & -10.11091(15.25728) & 80.91698(15.50048) \\ 
		\hline
		rate & 0.16626(0.26154) & 0.05287(0.2613) & 0.69135(0.43628) \\ 
		\hline
		ferrate & 2.58746***(0.96028) & -1.67848(1.06485) & 4.32561***(1.12471) \\ 
		powgend & -3.09196***(0.49015) & 0.714(0.41834) & -0.54734(0.51338) \\ 
		eq\_health & 0.10112(0.69408) & 2.75906***(0.78272) & 4.62427***(1.1305) \\ 
		ineq\_educ & -2.09696(2.33622) & 5.29634***(2.39962) & -10.3432***(2.70765) \\ 
		ineq\_inc & 5.32876***(2.59477) & -9.08242***(2.35931) & 10.89922***(3.40414) \\ 
		eq\_access & 28.56299***(5.5844) & -5.09848(6.35618) & 31.06497***(7.08197) \\ 
		eq\_redist & -40.91293***(12.89674) & -38.18629***(16.62724) & -114.90727***(19.31919) \\ 
		civilpart & -9.59618***(2.78319) & 5.08124***(2.47201) & -6.36655(4.05762) \\ 
		accjust & -1.89435(9.22897) & -5.28742(10.20244) & -75.66229***(12.17123) \\ 
		ecocont & 0.42103(0.29668) & -0.02385(0.3009) & -0.00473(0.41129) \\ 
		freexp & -15.37981**(7.11022) & 45.12384***(11.12141) & 46.82842***(10.41181) \\ 
		freerelig & 1.26604***(0.57401) & 0.68092(0.62653) & -0.2062(0.74503) \\ 
		lifexp& 2.02242(2.7095) & 6.54206***(2.64635) & 12.97523***(5.51876) \\ 
		gdppc & -0.8286***(0.21716) & -0.14708(0.17659) & -0.15251(0.2456) \\ 
		\hline
		\textit{Note:}  & \multicolumn{3}{r}{$^{*}$p$<$0.1; $^{**}$p$<$0.05; $^{***}$p$<$0.01} \\ 
		\specialrule{2pt}{0pt}{0pt}	
	\end{tabular} 
\end{table} 

\subsection{Model for G \label{gmodel}}

\begin{table}[H] \centering 
	\scriptsize
	\caption{Models for the neighborhood treatment $\boldsymbol{G}_{i}$: multivariate linear model \label{tab: g}}
	\renewcommand*{\arraystretch}{1.5}
	\begin{tabular}    {l@{\extracolsep{10pt}}l@{\extracolsep{10pt}}l@{\extracolsep{10pt}}l@{\extracolsep{10pt}} } 
		\specialrule{2pt}{0pt}{0pt}
		& \multicolumn{3}{c}{\textit{Dependent variable:}} \\ 
		\cline{2-4} 
		& \multicolumn{3}{c}{$\boldsymbol{G}^{*}$} \\ 
		& Statistic & Statistic & Statistic\\ 
		Omnibus Effect & 35.31*** & 38.27*** & 41.42*** \\ 
		(Intercept) & 21.63 ***& 28.17*** & 9.79*** \\ 
		\hline
		$Z_{i}$ & 12.09 ***& 14.62*** & 9.81*** \\ 
		\hline
		rate & 2.71 **& 4.80 **& 0.91 \\ 
		\hline
		ferrate & 18.01 ***& 20.04 ***& 14.95 ***\\ 
		powgend & 3.76 ***& 6.22 ***& 3.44 ***\\ 
		eq\_health & 6.61 *** & 12.35*** & 2.34 ***\\ 
		ineq\_educ & 0.38 & 0.07 & 5.43***\\ 
		ineq\_inc & 11.17 *** & 11.28 ***& 8.85***\\ 
		eq\_access & 7.81***& 12.64***& 3.32 ***\\ 
		eq\_resdist & 15.75 ***& 24.51***& 4.97*** \\ 
		civilpart & 16.13 ***& 18.06***& 14.58 ***\\ 
		accjust & 20.43 ***& 21.78 ***& 12.37\\ 
		econcont & 16.66 ***& 16.39 *** & 20.85 ***\\ 
		freeexp & 15.85 ***& 16.85 ***& 9.86 ***\\ 
		freerelig & 14.43 ***& 13.58 ***& 18.86 ***\\ 
		lifeexp & 1.01 & 0.64 & 4.51 ***\\ 
		gdppc & 191.40 ***& 196.56*** & 180.59 ***\\ 
		Vertex\_centr & 72.79 ***& 81.82 ***& 124.79 *** \\ 
		\specialrule{1.5pt}{0pt}{0pt}		
		\text{IIW} & & &  \\
		$\alpha$ & 1/2 & 1 & 0   \\
		$\beta$ & 1/2 & 0 & 1  \\
		\hline
		\textit{Note:}  & \multicolumn{3}{r}{$^{*}$p$<$0.1; $^{**}$p$<$0.05; $^{***}$p$<$0.01} \\ 
		\specialrule{2pt}{0pt}{0pt}	
	\end{tabular} 
\end{table} 

\subsection{Models for Y \label{ymodel}}
\begin{table}[H] \centering 
	\scriptsize
	\label{tab: y}
	\caption{Models for Y: linear model with time fixed effects} 
	\renewcommand*{\arraystretch}{1}
	\setlength{\tabcolsep}{1pt}
	\label{} 
	\centering
	\renewcommand*{\arraystretch}{1.5}
	\begin{tabular}{l@{\extracolsep{10pt}}l@{\extracolsep{10pt}}l@{\extracolsep{10pt}}l@{\extracolsep{10pt}}}
		\specialrule{2pt}{0pt}{0pt}
		& \multicolumn{3}{c}{\textit{Dependent variable:}} \\ 
		\cline{1-4} 
		& \multicolumn{3}{c}{Y} \\ 
		
		$Z_{i,HL}$  & 0.18591**(0.08543) & 0.04817(0.09693) & 0.1866**(0.07333) \\ 
		$Z_{i,LH}$  & 0.25183**(0.08896) & 0.20069**(0.08139) & 0.26622***(0.08028) \\ 
		$Z_{i,HH}$  & 0.20918*(0.11826) & 0.04082(0.08269) & 0.19946*(0.11675) \\ 
		\hline
		$G^{*}_{i,LL}$ & -0.44931(0.32531) & 0.46268(0.32837) & -0.92524***(0.28568) \\ 
		$G^{*}_{i,HL}$ & -0.26138(0.35755) & -1.4981**(0.58176) & -0.00591(0.22521) \\ 
		$G^{*}_{i,LH}$ & 0.9401**(0.47619) & 1.80662***(0.53027) & 0.33123(0.3205) \\ 
		$G^{*}_{i,HH}$ & 1.46947***(0.36545) & 0.55599(0.36047) & 1.05639***(0.3828) \\ 
		\hline
		
		$\phi(z_{i};\boldsymbol{X}^{z}_{i})$  & -0.07777(0.06924) & -0.14616**(0.07232) & -0.08396(0.0706) \\ 
		$\lambda(\boldsymbol{g}_{i};z_{i},\boldsymbol{X}^{g}_{i})$  & 0.33052(**0.16081) & 0.36958**(0.15836) & 0.05239(0.14535) \\ 
		\hline
		\text{IIW} & & &  \\
		$\alpha$ & 1/2 & 1 & 0   \\
		$\beta$ & 1/2 & 0 & 1   \\
		\hline
		\textit{Note:}  & \multicolumn{3}{r}{$^{*}$p$<$0.1; $^{**}$p$<$0.05; $^{***}$p$<$0.01} \\ 
		\specialrule{2pt}{0pt}{0pt}	
	\end{tabular}
\end{table}

\end{document}